\documentstyle[aps,version2,preprint]{revtex}    % preprint form
\begin{document}

\draft

\begin{title}
Electrons, pseudoparticles, and quasiparticles in the\\
one-dimensional many-electron problem                      
\end{title}

\author{J. M. P. Carmelo$^{1}$ and A. H. Castro Neto$^{2}$} 
\begin{instit}
$^{1}$ Department of Physics, University of \'Evora,
Apartado 94, P-7001 \'Evora Codex, Portugal\\
and Centro de F\'{\i}sica das Interac\c{c}\~oes Fundamentais,  
I.S.T., P-1096 Lisboa Codex, Portugal
\end{instit}
\begin{instit}
$^{2}$ Department of Physics, University of California,
Riverside, CA 92521
\end{instit}
\receipt{February 1996}

\begin{abstract}
We generalize the concept of quasiparticle for one-dimensional 
(1D) interacting electronic systems. The $\uparrow $ and $\downarrow $ 
quasiparticles recombine the pseudoparticle colors $c$ and $s$ 
(charge and spin at zero magnetic field) and are constituted
by one many-pseudoparticle {\it topological momenton} and one or 
two pseudoparticles. These excitations cannot be separated. We 
consider the case of the Hubbard chain. We show that the low-energy 
electron -- quasiparticle transformation has a singular charater 
which justifies the perturbative and non-perturbative nature of 
the quantum problem in the pseudoparticle and electronic 
basis, respectively. This follows from the absence of 
zero-energy electron -- quasiparticle overlap in 1D. 
The existence of Fermi-surface quasiparticles both in 1D and 
three dimensional (3D) many-electron systems suggests there 
existence in quantum liquids in dimensions 1$<$D$<$3. However,
whether the electron -- quasiparticle overlap can vanish in D$>$1 
or whether it becomes finite as soon as we leave 1D remains an 
unsolved question.
\end{abstract}
\renewcommand{\baselinestretch}{1.656}   % for preprint style only


\narrowtext

%%%%%%%%%%%%%%%%%%%%%%%%%%%%%%%%%%%%%%%%%%%%%%%%%%%%%%%%%%%%%%%%
\section{INTRODUCTION}

The unconventional electronic properties of novel materials 
such as the superconducting coper oxides and synthetic 
quasi-unidimensional conductors has attracted much attention to 
the many-electron problem in spatial dimensions 1$\leq$D$\leq$3. 
A good understanding of {\it both} the different and common 
properties of the 1D and 3D many-electron problems might 
provide useful indirect information on quantum liquids in dimensions  
1$<$D$<$3. This is important because the direct study of the 
many-electron problem in dimensions 1$<$D$<$3 is of great 
complexity. The nature of interacting electronic quantum liquids in 
dimensions 1$<$D$<$3, including the existence or non 
existence of quasiparticles and Fermi surfaces, remains an 
open question of crucial importance for the clarification of
the microscopic mechanisms behind the unconventional
properties of the novel materials. 

In 3D the many-electron quantum problem can often be described 
in terms of a one-particle quantum problem of quasiparticles 
\cite{Pines,Baym}, which interact only weakly. This Fermi liquid 
of quasiparticles describes successfully the properties of most 
3D metals, which are not very sensitive to the presence of 
electron-electron interactions. There is a one to one
correspondence between the $\sigma $ quasiparticles and the 
$\sigma $ electrons of the original non-interacting problem
(with $\sigma =\uparrow \, , \downarrow$). 
Moreover, the coherent part of the $\sigma $ one-electron Green 
function is quite similar to a non-interacting Green function 
except that the bare $\sigma $ electron spectrum is replaced by the
$\sigma $ quasiparticle spectrum and for an electron renormalization
factor, $Z_{\sigma }$, smaller than one and such that
$0<Z_{\sigma }<1$. A central point of Fermi-liquid theory
is that quasiparticle - quasihole processes describe exact 
low-energy and small-momentum Hamiltonian eigenstates and ``adding'' or 
``removal'' of one quasiparticle connects two exact ground states 
of the many-electron Hamiltonian. 

On the other hand, in 1D many-electron systems 
\cite{Solyom,Haldane,Metzner}, such as the hUBbard chain
solvable by Bethe ansatz (BA) \cite{Bethe,Yang,Lieb,Korepinrev}, 
the $\sigma $ electron renormalization factor, $Z_{\sigma }$, 
vanishes \cite{Anderson,Carmelo95a}. Therefore, the 
many-particle problem is not expected to be descibed in terms of a 
one-particle problem of Fermi-liquid quasiparticles. Such 
non-perturbative electronic problems are usually called Luttinger 
liquids \cite{Haldane}. In these systems the two-electron vertex 
function at the Fermi momentum diverges in the limit of 
vanishing excitation energy \cite{Carmelo95a}. 
In a 3D Fermi liquid this quantity is closely related to the 
interactions of the quasiparticles \cite{Pines,Baym}. 
Its divergence seems to indicate that there are no quasiparticles 
in 1D interacting electronic systems. A second possibility 
is that there are quasiparticles in the 1D many-electron
problem but without overlap with the electrons in the limit 
of vanishing excitation energy.

While the different properties of 1D and 3D many-electron
problems were the subject of many Luttinger-liquid studies 
in 1D \cite{Solyom,Haldane,Metzner}, the characterization of 
their common properties is also of great interest because the 
latter are expected to be present in dimensions 1$<$D$<$3 as 
well. One example is the Landau-liquid character common to 
Fermi liquids and some Luttinger liquids which consists in the 
generation of the low-energy excitations in terms of different 
momentum-occupation configurations of anti-commuting quantum objects 
(quasiparticles or pseudoparticles) whose forward-scattering 
interactions determine the low-energy properties of the 
quantum liquid. This generalized Landau-liquid theory was 
first applied in 1D to contact-interaction soluble problems 
\cite{Carmelo90} and shortly after also to $1/r^2$-interaction 
integrable models \cite{Haldane91}. Within this picture
the 1D many-electron problem can also be described in
terms of weakly interacting ``one-particle'' objects,
the pseudoparticles, which, however, have no one-to-one
correspondence with the electrons, as is shown in this paper.

In spite of the absence of the one to one
principle in what concerns single pseudoparticles
and single electrons, following the studies of
Refs. \cite{Carmelo90,Carmelo91b,Carmelo92}
a generalized adiabatic principle for small-momentum
pseudoparticle-pseudohole and electron-hole excitations
was introduced for 1D many-electron problems
in Refs. \cite{Carmelo92b}.
The pseudoparticles of 1D many-electron systems show other
similarities with the quasiparticles of a Fermi liquid, there 
interactions being determined by {\it finite} forward-scattering
$f$ functions \cite{Carmelo91b,Carmelo92,Carmelo92b}.
At constant values of the electron numbers this description of 
the quantum problem is very similar to Fermi-liquid theory, 
except for two main differences: (i) the $\uparrow $ and $\downarrow $ 
quasiparticles are replaced by the $c$ and $s$ pseudoparticles 
\cite{Carmelo93,Carmelo94,Carmelo94b,Carmelo94c,Carmelo95},
and (ii) the discrete pseudoparticle momentum (pseudomomentum) is 
of the usual form $q_j={2\pi\over {N_a}}I_j^{\alpha}$ 
but the numbers $I_j^{\alpha}$ (with $\alpha =c,s$) are not 
always integers. They are integers or half integers depending on
whether the number of particles in the system is even or odd.
This plays a central role in the present quasiparticle problem.
The connection of these perturbative pseudoparticles 
to the non-perturbative 1D electronic basis remains
an open problem. By perturbative we mean here the fact
that the two-pseudoparticle $f$ functions and forward-scattering
amplitudes are finite \cite{Carmelo92b,Carmelo94}, in contrast 
to the two-electron vertice functions.

The low-energy excitations of the Hubbard chain at constant electron 
numbers and in a finite magnetic field and chemical potential were shown
\cite{Carmelo91b,Carmelo92,Carmelo92b,Carmelo94,Carmelo94b,Carmelo94c}
to be $c$ and $s$ pseudoparticle-pseudohole processes
relative to the canonical-ensemble ground state. This determines 
the $c$ and $s$ low-energy separation \cite{Carmelo94c}, which at 
zero magnetization leads to the so called charge and spin separation.
In this paper we find that in addition to the above 
pseudoparticle-pseudohole excitations there are also Fermi-surface 
{\it quasiparticle} transitions in the 1D many-electron problem. 
Moreover, it is the study of such quasiparticle which clarifies 
the complex and open problem of the low-energy electron -- 
pseudoparticle transformation. 

As in 3D Fermi liquids, the 
quasiparticle excitation is a transition between two exact ground 
states of the interacting electronic problem differing in the number 
of electrons by one. When one electron is added to the electronic 
system the number of these excitations {\it also} increases
by one. Naturally, its relation to the electron excitation will 
depend on the overlap between the states associated with this and the 
quasiparticle excitation and how close we are in energy from the initial 
interacting ground state. Therefore, in order to define the 
quasiparticle we need to understand the properties of the actual 
ground state of the problem as, for instance, is given
by its exact solution via the BA. We find that in the 1D Hubbard
model adding one $\uparrow $ or $\downarrow $ electron of 
lowest energy is associated with adding one $\uparrow $ or
$\downarrow $ quasiparticle, as in a Fermi liquid. These are 
many-pseudoparticle objects which recombine the colors $c$ and $s$
giving rise to the spin projections $\uparrow $ and 
$\downarrow $. We find that the quasiparticle is constituted by
individual pseudoparticles and by a many-pseudoparticle object
of large momentum that we call topological momenton. Importantly,
these excitations cannot be separated. Although one quasiparticle is 
basically one electron, we show that in 1D the quasiparticle --
electron transformation is singular because it involves
the vanishing one-electron renormalization factor.
This also implies a low-energy singular electron - pseudoparticle 
transformation. This singular character explains why the problem 
becomes perturbative in the pseudoparticle basis while it is non 
perturbative in the usual electronic picture.

The singular nature of the low-energy electron - quasiparticle
and electron -- pseudoparticle transformations reflects the fact 
that the one-electron density of states vanishes in the 1D 
electronic problem when the excitation energy $\omega\rightarrow 0$.
The diagonalization of the many-electron problem is
at lowest excitation energy associated with the singular electron -- 
quasiparticle transformation which absorbes the vanishing electron 
renormalization factor and maps vanishing electronic spectral weight onto
finite quasiparticle and pseudoparticle spectral weight.
For instance, by absorbing the renormalization factor the 
electron - quasiparticle transformation renormalizes 
divergent two-electron vertex functions onto 
finite two-quasiparticle scattering parameters. These quantities 
fully determine the finite $f$ functions and scattering amplitudes
of the pseudoparticle theory \cite{Carmelo92,Carmelo92b,Carmelo94b}. 
The pseudoparticle $f$ functions and amplitudes determine all the 
static and low-energy quantities of the 1D many-electron problem 
and are associated with zero-momentum two-pseudoparticle 
forward scattering. 

The paper is organized as follows: the pseudoparticle
operator basis is summarized in Sec. II. In Sec. III we find
the quasiparticle operational expressions in the pseudoparticle
basis and characterize the corresponding $c$ and $s$
recombination in the $\uparrow $ and $\downarrow $ spin
projections. The singular electron -- quasiparticle (and
electron -- pseudoparticle) transformation is 
studied in Sec. IV. Finally, in Sec.
V we present the concluding remarks.

%%%%%%%%%%%%%%%%%%%%%%%%%%%%%%%%%%%%%%%%%%%%%%%%%%%%%%%%%%%%%%%%
\section{THE PERTURBATIVE PSEUDOPARTICLE OPERATOR BASIS}

It is useful for the studies presented in this paper to 
introduce in this section some basic information on
the perturbative operator pseudoparticle basis, as it is 
obtained directly from the BA solution 
\cite{Carmelo94,Carmelo94b,Carmelo94c}.
We consider the Hubbard model in 1D \cite{Lieb,Frahm,Frahm91}
with a finite chemical potential $\mu$ and in the
presence of a magnetic field $H$
\cite{Carmelo92b,Carmelo94,Carmelo94b}

\begin{eqnarray}
\hat{H} = -t\sum_{j,\sigma}\left[c_{j,\sigma}^{\dag }c_{j+1,\sigma}+c_
{j+1,\sigma}^{\dag }c_{j,\sigma}\right] +
U\sum_{j} [c_{j,\uparrow}^{\dag }c_{j,\uparrow} - 1/2]
[c_{j,\downarrow}^{\dag }c_{j,\downarrow} - 1/2]
- \mu \sum_{\sigma} \hat{N}_{\sigma } - 2\mu_0 H\hat{S}_z \, ,
\end{eqnarray}
where $c_{j,\sigma}^{\dag }$ and $c_{j,\sigma}$ are the creation and
annihilation operators, respectively, for electrons at the 
site $j$ with spin projection $\sigma=\uparrow, \downarrow$. 
In what follows $k_{F\sigma}=\pi n_{\sigma}$ and 
$k_F=[k_{F\uparrow}+k_{F\downarrow}]/2=\pi n/2$, where 
$n_{\sigma}=N_{\sigma}/N_a$ and $n=N/N_a$, and $N_{\sigma}$ and 
$N_a$ are the number of $\sigma$ electrons and lattice sites, 
respectively ($N=\sum_{\sigma}N_{\sigma}$). We also consider
the spin density, $m=n_{\uparrow}-n_{\downarrow}$.

The many-electron problem $(1)$ can be diagonalized using the 
BA \cite{Yang,Lieb}. We consider all finite values of $U$, 
electron densities $0<n<1$, and spin densities $0<m<n$. 
For this parameter space the low-energy physics is dominated 
by the lowest-weight states (LWS's) of the spin and eta-spin algebras 
\cite{Korepin,Essler} of type I \cite{Carmelo94,Carmelo94b,Carmelo95}. 
The LWS's I are described by real BA rapidities, whereas all
or some of the BA rapidities which describe the LWS's II 
are complex and non-real. Both the LWS's II and the non-LWS's 
out of the BA solution \cite{Korepin} have energy gaps relative 
to each canonical ensemble ground state 
\cite{Carmelo94,Carmelo94b,Carmelo95}. Fortunately, the
quasiparticle description involves only LWS's I because
these quantum objects are associated with ground-state
-- ground-state transitions and in the present parameter
space all ground states of
the model are LWS's I. On the other hand, the electronic
excitation involves transitions to LWS's I, LWS's II,
and non-LWS's, but the electron -- quasiparticle transformation
involves only LWS's I. Therefore, our results refer mainly 
to the Hilbert sub space spanned by the LWS's I and are
valid at energy scales smaller than the above gaps. 
(Note that in simpler 1D quantum problems of symmetry 
$U(1)$ the states I span the whole Hilbert space 
\cite{Anto}.)

In this Hilbert sub space the BA solution was shown to 
refer to an operator algebra which involves two types 
of {\it pseudoparticle} creation (annihilation) operators 
$b^{\dag }_{q,\alpha }$ ($b_{q,\alpha }$). These obey the 
usual anti-commuting algebra \cite{Carmelo94,Carmelo94b,Carmelo94c} 

\begin{equation}
\{b^{\dag }_{q,\alpha},b_{q',\alpha'}\} 
=\delta_{q,q'}\delta_{\alpha ,\alpha'}, \hspace{0.5cm} 
\{b^{\dag }_{q,\alpha},b^{\dag }_{q',\alpha'}\}=0, \hspace{0.5cm} 
\{b_{q,\alpha},b_{q',\alpha'}\}=0 \, .
\end{equation}       
Here $\alpha$ refers to the two pseudoparticle 
colors $c$ and $s$ \cite{Carmelo94,Carmelo94b,Carmelo94c}. The 
discrete pseudomomentum values are 

\begin{equation}
q_j = {2\pi\over {N_a}}I_j^{\alpha } \, , 
\end{equation}
where $I_j^{\alpha }$ are {\it consecutive} integers or half 
integers. There are $N_{\alpha }^*$ values of $I_j^{\alpha }$, {\it i.e.}
$j=1,...,N_{\alpha }^*$. A LWS I is specified by 
the distribution of $N_{\alpha }$ 
occupied values, which we call $\alpha $ pseudoparticles, over 
the $N_{\alpha }^*$ available values. There are $N_{\alpha }^*-
N_{\alpha }$ corresponding empty values, which we call $\alpha $ 
pseudoholes. These are good quantum numbers such that

\begin{equation}
N_c^* = N_a \, ; \hspace{0.5cm} 
N_c = N \, ; \hspace{0.5cm} 
N_s^* = N_{\uparrow} \, ; \hspace{0.5cm}
N_s = N_{\downarrow} \, . 
\end{equation}

The numbers $I_j^c$ are integers (or half integers) for 
$N_s$ even (or odd), and $I_j^s$ are integers (or 
half integers) for $N_s^*$ odd (or even) \cite{Lieb}. 
All the states I can be generated by acting onto
the vacuum $|V\rangle $ (zero-electron density) suitable
combinations of pseudoparticle operators 
\cite{Carmelo94,Carmelo94b}. The ground state 

\begin{equation}
|0;N_{\sigma }, N_{-\sigma}\rangle = \prod_{\alpha=c,s}
[\prod_{q=q_{F\alpha }^{(-)}}^{q_{F\alpha }^{(+)}} 
b^{\dag }_{q,\alpha }] 
|V\rangle \, ,
\end{equation}
and all LWS's I are Slatter determinants of pseudoparticle 
levels. In Appendix A we define the pseudo-Fermi points, 
$q_{F\alpha }^{(\pm )}$, of
$(5)$. In that Appendix we also present other quantities
of the pseudoparticle representation which are useful
for the present study.

In the pseudoparticle basis spanned by the LWS's I and 
in normal order relatively to the ground state $(5)$ the 
Hamiltonian $(1)$ has the following form \cite{Carmelo94,Carmelo94c}

\begin{equation}
:\hat{H}: = \sum_{i=1}^{\infty}\hat{H}^{(i)} \, ,
\end{equation} 
where, to second pseudoparticle scattering order

\begin{eqnarray}
\hat{H}^{(1)} & = & \sum_{q,\alpha} 
\epsilon_{\alpha}(q):\hat{N}_{\alpha}(q): \, ;\nonumber\\
\hat{H}^{(2)} & = & {1\over {N_a}}\sum_{q,\alpha} \sum_{q',\alpha'} 
{1\over 2}f_{\alpha\alpha'}(q,q') 
:\hat{N}_{\alpha}(q)::\hat{N}_{\alpha'}(q'): \, .
\end{eqnarray} 
Here $(7)$ are the Hamiltonian terms which are {\it 
relevant} at low energy \cite{Carmelo94b}. 
Furthermore, at low energy and small momentum the only relevant term is the
non-interacting term $\hat{H}^{(1)}$. Therefore,
the $c$ and $s$ pseudoparticles are non-interacting at the 
small-momentum and low-energy fixed point and the spectrum 
is described in terms of the bands $\epsilon_{\alpha}(q)$ 
(studied in detail in Ref. \cite{Carmelo91b}) in a pseudo-Brillouin 
zone which goes between $q_c^{(-)}\approx -\pi$ and $q_c^{(+)}\approx \pi$ 
for the $c$ pseudoparticles and $q_s^{(-)}\approx -k_{F\uparrow}$ 
and $q_s^{(+)}\approx k_{F\uparrow}$ for the $s$ 
pseudoparticles. In the ground state $(5)$ these are occupied
for $q_{F\alpha}^{(-)}\leq q\leq q_{F\alpha}^{(+)}$,
where the pseudo-Fermi points (A1)-(A3) are such that
$q_{Fc}^{(\pm)}\approx \pm 2k_F$ and
$q_{Fs}^{(\pm)}\approx \pm k_{F\downarrow}$
(see Appendix A).

At higher energies and (or ) large momenta the pseudoparticles 
start to interact via zero-momentum transfer forward-scattering 
processes of the Hamiltonian $(6)-(7)$. As in a Fermi liquid, 
these are associated with $f$ functions and Landau parameters 
\cite{Carmelo92,Carmelo94}, whose expressions we present in 
Appendix A, where we also present the expressions for simple 
pseudoparticle-pseudohole operators which are useful for the 
studies of next sections.

%%%%%%%%%%%%%%%%%%%%%%%%%%%%%%%%%%%%%%%%%%%%%%%%%%%%%%%%%%%%%%%%
\section{THE QUASIPARTICLES AND $c$ AND $s$ RECOMBINATION}

In this section we introduce the 1D quasiparticle and express
it in the pseudoparticle basis. In Sec. IV we find that this
clarifies the low-energy transformation between the electrons
and the pseudoparticles. We define the quasiparticle operator as
the generator of a ground-state -- ground-state transition.
The study of ground states of form $(5)$ differing in the number of 
$\sigma $ electrons by one reveals that their relative momentum
equals {\it presisely } the $U=0$ Fermi points, $\pm k_{F\sigma}$.
Following our definition, the quasiparticle operator, ${\tilde{c}}^{\dag 
}_{k_{F\sigma },\sigma }$, which creates one quasiparticle 
with spin projection $\sigma$ and momentum $k_{F\sigma}$  
is such that

\begin{equation}
{\tilde{c}}^{\dag }_{k_{F\sigma},\sigma}
|0; N_{\sigma}, N_{-\sigma}\rangle =
|0; N_{\sigma} + 1, N_{\sigma}\rangle \, .
\end{equation}
The quasiparticle operator defines a one-to-one correspondence
between the addition of one electron to the system and the creation
of one quasiparticle: the electronic excitation, 
$c^{\dag }_{k_{F\sigma},\sigma}|0; N_{\sigma}, N_{-\sigma}\rangle$, 
defined at the Fermi momentum but arbitrary energy, contains
a single quasiparticle, as we show in Sec. IV. In that
section we will study this excitation as we take the energy to be 
zero, that is, as we approach the Fermi surface, where the problem is
equivalent to Landau's.

Since we are discussing the problem of addition or removal of
one particle the boundary conditions play a crucial role.
As discussed in Secs. I and II, the available Hamiltonian
eigenstates I depend on the discrete numbers $I_j^{\alpha}$ 
of Eq. $(3)$ which can be integers of half-integers depending 
on whether the number of particles in the system is even or odd 
[the pseudomomentum is given by Eq. $(3)$]. When we add or 
remove one electron to or from the many-body system we have to 
consider the transitions between states with integer and
half-integer quantum numbers [or equivalently, between states 
with an odd (even) and even (odd) number of $\sigma $ 
electrons]. The transition between two ground states differing 
in the number of electrons by one is associated with two different 
processes: a backflow in the Hilbert space of the pseudoparticles 
with a shift of all the pseudomomenta by $\pm\frac{\pi}{N_a}$ 
[associated with the change from even (odd) to odd (even) number 
of particles], which we call {\it topological momenton},
and the creation of one or a pair of pseudoparticles at the 
pseudo-Fermi points. 

According to the integer or half-integer character of the 
$I_j^{\alpha}$ numbers we have four ``topological'' types of Hilbert 
sub spaces. Since that character depends on the parities of the 
electron numbers, we refer these sub spaces by the parities of 
$N_{\uparrow}$ and $N_{\downarrow}$, respectively: (a) even, even; 
(b) even, odd; (c) odd, even; and (d) odd, odd. The ground-state 
total momentum expression is different for each type of Hilbert sub 
space in such a way that the relative momentum, $\Delta P$, 
of $U>0$ ground states differing in $N_{\sigma }$ by one equals 
the $U=0$ Fermi points, ie $\Delta P=\pm k_{F\sigma }$. Moreover, 
we find that the above quasiparticle operator $\tilde{c}^{\dag 
}_{k_{F\sigma },\sigma }$ involves the generator of one low-energy 
and large-momentum topological momenton. The $\alpha $ 
topological momenton is associated with the backflow of the 
$\alpha $ pseudoparticle pseudomomentum band and cannot occur 
without a second type of excitation associated 
with the adding or removal of pseudoparticles. The 
$\alpha $-topological-momenton generator, $U^{\pm 1}_{\alpha }$,
is an unitary operator which controls the topological
transformations of the pseudoparticle Hamiltonian $(6)-(7)$.
For instance, in the $\Delta P=\pm k_{F\uparrow }$ transitions
(a)$\rightarrow $(c) and (b)$\rightarrow $(d) the 
Hamiltonian $(6)-(7)$ transforms as 

\begin{equation}
:H: \rightarrow U^{\pm\Delta N_{\uparrow}}_s
:H: U^{\mp\Delta N_{\uparrow}}_s
\, ,
\end{equation}
and in the $\Delta P=\pm k_{F\downarrow }$ transitions
(a)$\rightarrow $(b) and (c)$\rightarrow $(d) as 

\begin{equation}
:H:\rightarrow U^{\pm \Delta 
N_{\downarrow}}_c:H:U^{\mp \Delta N_{\downarrow}}_c
\, ,
\end{equation}
where $\Delta N_{\sigma}=\pm 1$ and the expressions 
of the generator $U^{\pm 1}_{\alpha }$ is obtained 
below.

In order to arrive to the expressions for the quasipaticle
operators and associate topological-momenton generators
$U^{\pm 1}_{\alpha }$ we refer again to the ground-state 
pseudoparticle representation $(5)$. For simplicity, we 
consider that the initial ground state of form $(5)$
is non degenerate and has zero momentum. Following
equations (A1)-(A3) this corresponds to the situation
when both $N_{\uparrow }$ and $N_{\downarrow }$ are
odd, ie the initial Hilbert sub space is of type (d). However, note 
that our results are independent of the choice of 
initial ground state. The pseudoparticle numbers of the 
initial state are $N_c=N_{\uparrow }+N_{\downarrow }$ and
$N_s=N_{\downarrow }$ and the pseudo-Fermi points 
$q_{F\alpha }^{(\pm)}$ are given in Eq. (A1).

We express the electronic and pseudoparticle numbers and 
pseudo-Fermi points of the final states in terms of the 
corresponding values for the initial state.
We consider here the case when the final ground state 
has numbers $N_{\uparrow }$ and $N_{\downarrow }+1$ and
momentum $k_{F\downarrow }$. The procedures for final states
with these numbers and momentum $-k_{F\downarrow }$ or
numbers $N_{\uparrow }+1$ and $N_{\downarrow }$ and
momenta $\pm k_{F\uparrow }$ are similiar and are omitted here.

The above final state belongs the Hilbert sub space (c). 
Our goal is to find the quasiparticle operator 
$\tilde{c}^{\dag}_{k_{F\downarrow },\downarrow}$ such that

\begin{equation}
|0; N_{\uparrow}, N_{\downarrow}+1\rangle = 
\tilde{c}^{\dag}_{k_{F\downarrow },\downarrow} 
|0; N_{\uparrow}, N_{\downarrow}\rangle\, .
\end{equation}

Taking into account the changes in the pseudoparticle
quantum numbers associated with this (d)$\rightarrow $(c) 
transition we can write the final state as follows

\begin{equation}
|0; N_{\uparrow}, N_{\downarrow}+1\rangle = 
\prod_{q=q^{(-)}_{Fc}-\frac{\pi}{N_a}}^{q^{(+)}_{Fc}+
\frac{\pi}{N_a}}\prod_{q=q^{(-)}_{Fs}}^{q^{(+)}_{Fs}+\frac{2\pi}{N_a}}
b^{\dag}_{q,c} b^{\dag}_{q,s} |V\rangle \, ,
\end{equation}
which can be rewritten as

\begin{equation}
|0; N_{\uparrow}, N_{\downarrow}+1\rangle = 
b^{\dag}_{q^{(+)}_{Fc}+\frac{\pi}{N_a},c}
b^{\dag}_{q^{(+)}_{Fs}+\frac{2\pi}{N_a},s}
\prod_{q=q^{(-)}_{Fs}}^{q^{(+)}_{Fs}}
\prod_{q=q^{(-)}_{Fs}}^{q^{(+)}_{Fs}}
b^{\dag}_{q-\frac{\pi}{N_a},c} b^{\dag}_{q,s} |V\rangle \, ,
\end{equation}
and further, as

\begin{equation}
|0; N_{\uparrow}, N_{\downarrow}+1\rangle = 
b^{\dag}_{q^{(+)}_{Fc}+\frac{\pi}{N_a},c}
b^{\dag}_{q^{(+)}_{Fs}+\frac{2\pi}{N_a},s}
U_c^{+1}|0; N_{\uparrow}, N_{\downarrow}\rangle \, ,
\end{equation}
where $U_c^{+1}$ is the generator of expression $(10)$. Both
this operator and the operator $U_s^{+1}$ of Eq. $(9)$
obey the relation

\begin{equation}
U^{\pm 1}_{\alpha }b^{\dag }_{q,\alpha }U^{\mp 1}_{\alpha }= 
b^{\dag }_{q\mp {\pi\over {N_a}},\alpha }
\, .
\end{equation}
The pseudoparticle vacuum remains invariant under the
application of $U^{\pm 1}_{\alpha }$

\begin{equation}
U^{\pm 1}_{\alpha }|V\rangle = |V\rangle \, . 
\end{equation}
(The $s$-topological-momenton generator, $U_s^{+1}$, appears if we 
consider the corresponding expressions for the up-spin electron.) 
Note that the $\alpha $ topological momenton is an excitation 
which only changes the integer or half-integer character of the 
corresponding pseudoparticle quantum numbers $I_j^{\alpha }$. 
In Appendix B we derive the following expression for
the generator $U^{\pm 1}_{\alpha }$ 

\begin{equation}
U^{\pm 1}_{\alpha }=U_{\alpha } 
\left(\pm\frac{\pi}{N_a}\right)
\, ,
\end{equation}
where

\begin{equation}
U_{\alpha}(\delta q) = \exp\left\{ - i\delta q 
G_{\alpha}\right\} \, ,
\end{equation}
and

\begin{equation}
G_{\alpha} = -i\sum_{q} \left[{\partial\over 
{\partial q}} b^{\dag }_{q,\alpha }\right]b_{q,\alpha }
\, ,
\end{equation}
is the Hermitian generator of the $\mp {\pi\over {N_a}}$
topological $\alpha $ pseudomomentum translation. The operator 
$U^{\pm 1}_{\alpha }$ has the following discrete
representation

\begin{equation}
U^{\pm 1}_{\alpha } = \exp\left\{\sum_{q} 
b^{\dag }_{q\pm {\pi\over {N_a}},\alpha }b_{q,\alpha }\right\}
\, .
\end{equation}
When acting on the initial ground state of form $(5)$ the 
operator $U^{\pm 1}_{\alpha }$ produces a vanishing-energy 
$\alpha $ topological momenton of large momentum, $k=\mp N_{\alpha 
}{\pi\over {N_a}}\simeq q_{F\alpha}^{(\mp)}$. As referred above, 
the topological momenton is always combined with adding or 
removal of pseudoparticles.

In the two following equations we change notation and use 
$q_{F\alpha }^{(\pm)}$ to refer the pseudo-Fermi points of 
the final state (otherwise our reference state is the
initial state). Comparing equations $(11)$ and $(14)$ it 
follows that

\begin{equation}
\tilde{c}^{\dag }_{\pm k_{F\downarrow },\downarrow } =
b^{\dag }_{q_{Fc}^{(\pm)},c}b^{\dag }_{q_{Fs}^{(\pm)},s}
U^{\pm 1}_{c } \, ,
\end{equation}
and a similar procedure for the up-spin electron
leads to

\begin{equation}
\tilde{c}^{\dag }_{\pm k_{F\uparrow },\uparrow } =
b^{\dag }_{q_{Fc}^{(\pm)},c} 
U^{\pm 1}_{s} \, .
\end{equation}
According to these equations the $\sigma $ quasiparticles are 
constituted by one topological momenton and one or two
pseudoparticles. The topological momenton cannot be separated
from the pseudoparticle excitation, ie both these excitations
are confined inside the quasiparticle. Moreover, since the
generators $(17)-(20)$ have a many-pseudoparticle character,
following Eqs. $(21)-(22)$ the quasiparticle is a many-pseudoparticle 
object. Note also that both the $\downarrow $ and $\uparrow $ quasiparticles 
$(21)$ and $(22)$, respectively, are constituted by $c$ and $s$ 
excitations. Therefore, the $\sigma $ quasiparticle is a quantum object
which recombines the pseudoparticle colors $c$ and $s$ 
(charge and spin in the limit $m\rightarrow 0$ \cite{Carmelo94}) 
giving rise to spin projection $\uparrow $ or $\downarrow $.
It has ``Fermi surface'' at $\pm k_{F\sigma }$. 

However, two-quasiparticle objects can be of 
two-pseudoparticle character because the product of the 
two corresponding many-pseudoparticle operators is such 
that $U^{+ 1}_{\alpha }U^{- 1}_{\alpha }=\openone$, as for 
the triplet pair $\tilde{c}^{\dag }_{+k_{F\uparrow },\uparrow }
\tilde{c}^{\dag }_{-k_{F\uparrow },\uparrow }=
b^{\dag }_{q_{Fc}^{(+)},c}b^{\dag }_{q_{Fc}^{(-)},c}$.
Such triplet quasiparticle pair is constituted
only by individual pseudoparticles because it involves the mutual
annihilation of the two topological momentons
of generators $U^{+ 1}_{\alpha }$ and $U^{- 1}_{\alpha }$.
Therefore, relations $(21)$ and $(22)$ which connect quasiparticles
and pseudoparticles have some similarities with the
Jordan-Wigner transformation. 

Finally, we emphasize that the Hamiltonian-eigenstate generators 
of Eqs. $(26)$ and $(27)$ of Ref. \cite{Carmelo94b} are not 
general and refer to finite densities of added and removed 
electrons, respectively, corresponding to even electron 
numbers. The corresponding general generator expressions will
be studied elsewhere and involve the topological-momenton 
generators $(17)-(20)$.

%%%%%%%%%%%%%%%%%%%%%%%%%%%%%%%%%%%%%%%%%%%%%%%%%%%%%%%%%%%%%%%%
\section{THE ELECTRON - QUASIPARTICLE TRANSFORMATION}

In this section we study the relation of the 1D quasiparticle 
introduced in Sec. III to the electron. This study brings about
the question of the low-excitation-energy relation between the 
electronic operators $c_{k,\sigma}^{\dag }$ in momentum space at 
$k=\pm k_{F\sigma }$ and the pseudoparticle operators 
$b_{q,\alpha}^{\dag }$ at the pseudo-Fermi points. 

The quasiparticle operator, ${\tilde{c}}^{\dag }_{k_{F\sigma 
},\sigma}$, which creates one quasiparticle with spin projection 
$\sigma$ and momentum $k_{F\sigma}$, is defined by Eq.
$(8)$. In the pseudoparticle basis the $\sigma $ quasiparticle 
operator has the form $(21)$ or $(22)$. However, since we do not
know the relation between the electron and the pseudoparticles,
Eqs. $(21)$ and $(22)$ do not provide direct information
on the electron content of the $\sigma $ quasiparticle. Equation
$(8)$ tells us that the quasiparticle operator defines a one-to-one 
correspondence between the addition of one electron to the system 
and the creation of one quasiparticle, exactly as we expect 
from the Landau theory in 3D: the electronic excitation, 
$c^{\dag }_{k_{F\sigma },\sigma}|0; N_{\uparrow}=N_c-N_s,  
N_{\downarrow}=N_s\rangle$, defined at the 
Fermi momentum but arbitrary energy, contains a single $\sigma $ 
quasiparticle, as we show below. When we add or remove one 
electron from the many-body system this includes the transition
to the suitable final ground state as well as transitions
to excited states. The former transition is nothing but
the quasiparticle excitation of Sec. III.

Although our final results refer to momenta $k=\pm k_{F\sigma }$,
in the following analysis we consider for simplicity only the 
momentum $k=k_{F\sigma }$. In order to relate the quasiparticle 
operators $\tilde{c}^{\dag }_{k_{F\sigma },\sigma }$ to
the electronic operators $c^{\dag }_{k_{F\sigma },\sigma }$ 
we start by defining the Hilbert sub space 
where the low-energy $\omega $ projection of the state

\begin{equation}
c^{\dag }_{k_{F\sigma},\sigma}
|0; N_{\sigma}, N_{-\sigma} \rangle \, , 
\end{equation}
is contained. Notice that the electron excitation $(23)$ is {\it 
not} an eigenstate of the interacting problem:
when acting onto the initial ground state
$|0;i\rangle\equiv |0; N_{\sigma}, N_{-\sigma} \rangle$
the electronic operator
$c^{\dag }_{k_{F\sigma},\sigma }$ can be written as

\begin{equation}
c^{\dag }_{k_{F\sigma},\sigma } = \left[\langle 0;f|c^{\dag 
}_{k_{F\sigma},\sigma }|0;i\rangle + {\hat{R}}\right]
\tilde{c}^{\dag }_{k_{F\sigma},\sigma } \, ,
\end{equation}
where 

\begin{equation}
{\hat{R}}=\sum_{\gamma}\langle \gamma;k=0|c^{\dag 
}_{k_{F\sigma},\sigma }|0;i\rangle {\hat{A}}_{\gamma} \, , 
\end{equation}
and 

\begin{equation}
|\gamma;k=0\rangle = {\hat{A}}_{\gamma}
\tilde{c}^{\dag }_{k_{F\sigma },\sigma }|0;i\rangle
= {\hat{A}}_{\gamma}|0;f\rangle \, .
\end{equation}
Here $|0;f\rangle\equiv |0; N_{\sigma}+1, N_{-\sigma} \rangle$
denotes the final ground state, $\gamma$ represents the set of 
quantum numbers needed to specify each Hamiltonian eigenstate
present in the excitation $(23)$, 
and ${\hat{A}}_{\gamma}$ is the corresponding generator. The 
first term of the rhs of Eq. $(24)$ refers to the ground state - ground 
state transition and the operator $\hat{R}$ generates $k=0$ 
transitions from $|0,f\rangle $ to states I, states II, 
and non LWS's. Therefore, the electron excitation $(23)$
contains the quantum superposition of both the suitable final 
ground state $|0;f\rangle$, of excited states I relative to that 
state which result from multiple pseudoparticle-pseudohole processes 
associated with transitions to states I, and of 
LWS's II and non-LWS's. All these states have the 
same electron numbers as the final ground state. The transitions 
to LWS's II and to non-LWS's require a minimal finite energy 
which equals their gap relative to the final ground state.
The set of all these Hamiltonian eigenstates spans the
Hilbert sub space where the electronic operators $c^{\dag }_{k_{F\sigma 
},\sigma }$ $(24)$ projects the initial ground state. 

In order to show that the ground-state -- ground-state leading order 
term of $(24)$ controls the low-energy physics, we study the low-energy 
sector of the above Hilbert sub space. This is spanned by low-energy 
states I. In the case of these states the generator 
${\hat{A}}_{\gamma}$ of Eq. $(26)$ reads

\begin{equation}
{\hat{A}}_{\gamma}\equiv 
{\hat{A}}_{\{N_{ph}^{\alpha ,\iota}\},l} = \prod_{\alpha=c,s}
{\hat{L}}^{\alpha\iota}_{-N_{ph}^{\alpha\iota}}(l) \, , 
\end{equation}
where the operator ${\hat{L}}^{\alpha\iota}_{-N_{ph}^{\alpha\iota}}(l)$
is given in Eq. $(56)$ of Ref. \cite{Carmelo94b} and
produces a number $N_{ph}^{\alpha ,\iota}$ of 
$\alpha ,\iota$ pseudoparticle-pseudohole
processes onto the final ground state. Here $\iota =sgn (q)1=\pm 1$ 
defines the right ($\iota=1$) and left ($\iota=-1$) 
pseudoparticle movers, $\{N_{ph}^{\alpha ,\iota}\}$ 
is a short notation for

\begin{equation}
\{N_{ph}^{\alpha ,\iota}\}\equiv 
N_{ph}^{c,+1}, N_{ph}^{c,-1}, 
N_{ph}^{s,+1}, N_{ph}^{s,-1} \, , 
\end{equation}
and $l$ is a quantum number which distinguishes 
different pseudoparticle-pseudohole distributions 
characterized by the same values for the numbers $(28)$.
In the case of the lowest-energy states I
the above set of quantum numbers $\gamma $ is thus 
given by $\gamma\equiv \{N_{ph}^{\alpha ,\iota}\},l$.
(We have introduced the argument $(l)$ in the
operator $L^{\alpha\iota}_{-N_{ph}^{\alpha\iota}}(l)$
which for the same value of the $N_{ph}^{\alpha\iota}$ 
number defines different $\alpha\iota$ pseudoparticle - 
pseudohole configurations associated with different
choices of the pseudomomenta in the summation of 
expression $(56)$ of Ref. \cite{Carmelo94b}.) 
In the particular case of the lowest-energy states
expression $(26)$ reads

\begin{equation}
|\{N_{ph}^{\alpha ,\iota}\},l;k=0\rangle = 
{\hat{A}}_{\{N_{ph}^{\alpha ,\iota}\},l}
\tilde{c}^{\dag }_{k_{F\sigma },\sigma }|0;i\rangle
= {\hat{A}}_{\{N_{ph}^{\alpha ,\iota}\},l}|0;f\rangle \, .
\end{equation}
The full electron -- quasiparticle transformation $(24)$ involves 
other Hamiltonian eigenstates which are irrelevant for the 
quasiparticle problem studied in the present paper.
Therefore, we omit here the study of the general generators 
${\hat{A}}_{\gamma}$ of Eq. $(26)$. 

The momentum expression (relative to the final ground 
state) of Hamiltonian eigenstates with generators of
the general form $(27)$ is \cite{Carmelo94b}

\begin{equation}
k = {2\pi\over {N_a}}\sum_{\alpha ,\iota}\iota
N_{ph}^{\alpha\iota} \, . 
\end{equation}
Since our states $|\{N_{ph}^{\alpha ,\iota}\},l;k=0\rangle$ 
have zero momentum relative to the final ground state 
they have restrictions in the choice of the numbers $(28)$. 
For these states these numbers are such that

\begin{equation}
\sum_{\alpha ,\iota}\iota
N_{ph}^{\alpha ,\iota} = 0 \, , 
\end{equation}
which implies that

\begin{equation}
\sum_{\alpha }N_{ph}^{\alpha ,+1} = 
\sum_{\alpha }N_{ph}^{\alpha ,-1} =
\sum_{\alpha }N_{ph}^{\alpha ,\iota} \, . 
\end{equation}
Since 

\begin{equation}
N_{ph}^{\alpha ,\iota}=1,2,3,.... \, ,
\end{equation}
it follows from Eqs. $(31)-(33)$ that

\begin{equation}
\sum_{\alpha ,\iota}
N_{ph}^{\alpha ,\iota} = 2,4,6,8.... \, , 
\end{equation}
is always an even positive integer.

The vanishing chemical-potential excitation energy,

\begin{equation}
\omega^0_{\sigma }=\mu(N_{\sigma }+1,N_{-\sigma })
-\mu(N_{\sigma },N_{-\sigma }) \, ,
\end{equation}
can be evaluated by use of the Hamiltonian $(6)-(7)$ and is 
given by

\begin{equation}
\omega^0_{\uparrow } = {\pi\over {2N_a}}\left[v_c + F_{cc}^1
+ v_s + F_{ss}^1 - 2F_{cs}^1 + v_c + F_{cc}^0\right] \, ,
\end{equation}
and 

\begin{equation}
\omega^0_{\downarrow } = {\pi\over {2N_a}}\left[v_s + F_{ss}^1 
+ v_c + F_{cc}^0 + v_s + F_{ss}^0 + 2F_{cs}^0\right] \, ,
\end{equation}
for up and down spin, respectively, and involves the pseudoparticle 
velocities (A6) and Landau parameters (A8). Since we
measure the chemical potencial from its value at
the canonical ensemble of the reference initial ground state, 
ie we consider $\mu(N_{\sigma },N_{-\sigma })=0$,
$\omega^0_{\sigma }$ measures also the ground-state
excitation energy $\omega^0_{\sigma }=E_0(N_{\sigma 
}+1,N_{-\sigma })-E_0(N_{\sigma },N_{-\sigma })$.

The excitation energies $\omega (\{N_{ph}^{\alpha ,\iota}\})$ 
of the states $|\{N_{ph}^{\alpha ,\iota}\},l;k=0\rangle$ 
(relative to the initial ground state) involve the energy 
$\omega^0_{\sigma }$ and are $l$ independent. They are
given by

\begin{equation}
\omega (\{N_{ph}^{\alpha ,\iota}\}) =
\omega^0_{\sigma } + {2\pi\over {N_a}}\sum_{\alpha ,\iota}
v_{\alpha} N_{ph}^{\alpha ,\iota} \, . 
\end{equation}
We denote by $N_{\{N_{ph}^{\alpha ,\iota}\}}$ the number
of these states which obey the condition-equations 
$(31)$, $(32)$, and $(34)$ and have the same values for 
the numbers $(28)$. 

In order to study the main corrections to the
(quasiparticle) ground-state -- ground-state transition 
it is useful to consider the simplest case when $\sum_{\alpha 
,\iota}N_{ph}^{\alpha ,\iota}=2$. In this case we have 
$N_{\{N_{ph}^{\alpha ,\iota}\}}=1$ and, therefore, we can 
omit the index $l$. There are four of such Hamiltonian eigenstates. 
Using the notation of the right-hand side (rhs) of Eq. $(28)$
these states are $|1,1,0,0;k=0\rangle $, $|0,0,1,1;k=0\rangle $,
$|1,0,0,1;k=0\rangle $, and $|0,1,1,0;k=0\rangle $. They
involve two pseudoparticle-pseudohole processes with $\iota=1$ and 
$\iota=-1$, respectively and read

\begin{equation}
|1,1,0,0;k=0\rangle = \prod_{\iota=\pm 1}
{\hat{\rho}}_{c ,\iota}(\iota {2\pi\over {N_a}})
\tilde{c}^{\dag }_{k_{F\sigma },\sigma }
|0;i\rangle \, , 
\end{equation}

\begin{equation}
|0,0,1,1;k=0\rangle = \prod_{\iota=\pm 1}
{\hat{\rho}}_{s ,\iota}(\iota {2\pi\over {N_a}})
\tilde{c}^{\dag }_{k_{F\sigma },\sigma }
|0;i\rangle \, , 
\end{equation}

\begin{equation}
|1,0,0,1;k=0\rangle = 
{\hat{\rho}}_{c,+1}({2\pi\over {N_a}})
{\hat{\rho}}_{s,-1}(-{2\pi\over {N_a}})
\tilde{c}^{\dag }_{k_{F\sigma },\sigma }
|0;i\rangle \, , 
\end{equation}

\begin{equation}
|0,1,1,0;k=0\rangle = 
{\hat{\rho}}_{c,-1}(-{2\pi\over {N_a}})
{\hat{\rho}}_{s,+1}({2\pi\over {N_a}})
\tilde{c}^{\dag }_{k_{F\sigma },\sigma }
|0;i\rangle \, , 
\end{equation}
where ${\hat{\rho}}_{\alpha ,\iota}(k)$ is
the fluctuation operator of Eq. (A12). This was
studied in some detail in Ref. \cite{Carmelo94c}.

From equations $(26)$, $(27)$, and $(29)$ we can rewrite 
expression $(24)$ as

\begin{eqnarray}
c^{\dag }_{k_{F\sigma},\sigma } & = & 
\langle 0;f|c^{\dag }_{k_{F\sigma},\sigma }|0;i\rangle 
\left[1 + \sum_{\{N_{ph}^{\alpha ,\iota}\},l}
{\langle \{N_{ph}^{\alpha ,\iota}\},l;k=0|
c^{\dag }_{k_{F\sigma},\sigma}|0;i\rangle \over \langle 
0;f|c^{\dag }_{k_{F\sigma},\sigma }|0;i\rangle} \prod_{\alpha=c,s}
{\hat{L}}^{\alpha\iota}_{-N_{ph}^{\alpha\iota}}(l)\right]
\tilde{c}^{\dag }_{k_{F\sigma},\sigma } 
\nonumber\\
& + & \sum_{\gamma '}\langle \gamma ';k=0|c^{\dag 
}_{k_{F\sigma},\sigma }|0;i\rangle {\hat{A}}_{\gamma '}
\tilde{c}^{\dag }_{k_{F\sigma},\sigma } \, ,
\end{eqnarray}
where $\gamma '$ refers to the Hamiltonian eigenstates
of form $(26)$ whose generator ${\hat{A}}_{\gamma '}$
are not of the particular form $(27)$.

In Appendix C we evaluate the matrix elements of 
expression $(43)$ corresponding to transitions to the
final ground state and excited states of form $(29)$. 
Following Ref. \cite{Carmelo94b}, these states refer to the 
conformal-field-theory \cite{Frahm,Frahm91} critical point. 
They are such that the ratio $N_{ph}^{\alpha ,\iota}/N_a$ 
vanishes in the thermodynamic limit, $N_a\rightarrow 0$.
Therefore, in that limit the positive excitation
energies $\omega (\{N_{ph}^{\alpha ,\iota}\})$ of Eq. $(38)$
are vanishing small. The results of that Appendix lead to

\begin{equation}
\langle 0;f|c^{\dag }_{k_{F\sigma},\sigma}|0;i\rangle 
= \sqrt{Z_{\sigma }}\, , 
\end{equation}
where, as in a Fermi liquid \cite{Nozieres}, the one-electron 
renormalization factor

\begin{equation}
Z_{\sigma}=\lim_{\omega\to 0}Z_{\sigma}(\omega) \, ,
\end{equation}
is closed related to the $\sigma $ self energy $\Sigma_{\sigma} 
(k,\omega)$. Here the function $Z_{\sigma}(\omega)$ is given by 
the small-$\omega $ leading-order term of

\begin{equation}
|\varsigma_{\sigma}||1-{\partial \hbox{Re}
\Sigma_{\sigma} (\pm k_{F\sigma},\omega)
\over {\partial\omega}}|^{-1}\, ,
\end{equation}
where 

\begin{equation}
\varsigma_{\uparrow}=-2+\sum_{\alpha}
{1\over 2}[(\xi_{\alpha c}^1-\xi_{\alpha s}^1)^2
+(\xi_{\alpha c}^0)^2] \, , 
\end{equation}
and

\begin{equation}
\varsigma_{\downarrow}=-2 
+\sum_{\alpha}{1\over 2}[(\xi_{\alpha s}^1)^2+(\xi_{\alpha 
c}^0+\xi_{\alpha s}^0)^2] \, , 
\end{equation}
are $U$, $n$, and $m$ dependent exponents which for $U>0$ are 
negative and such that $-1<\varsigma_{\sigma}<-1/2$. In equations 
$(47)$ and $(48)$ $\xi_{\alpha\alpha'}^j$ are the parameters (A7). 
From equations $(46)$, (C11), and (C15) we find 

\begin{equation}
Z_{\sigma}(\omega)=a^{\sigma }_0
\omega^{1+\varsigma_{\sigma}} \, ,
\end{equation}
where $a^{\sigma }_0$ is a real and positive constant such
that 

\begin{equation}
\lim_{U\to 0}a^{\sigma }_0=1 \, .
\end{equation}

Equation $(49)$ confirms that the renormalization 
factor $(45)$ vanishes, as expected for a 1D many-electron 
problem \cite{Anderson}. It follows from Eq. $(44)$ that in 
the present 1D model the electron renormalization 
factor can be identified with a single matrix element 
\cite{Anderson,Metzner94}. We emphasize that in a Fermi liquid 
$\varsigma_{\sigma}=-1$ and Eq. $(46)$ recovers the usual 
Fermi-liquid relation. In the different three limits 
$U\rightarrow 0$, $m\rightarrow 0$, and $m\rightarrow n$ the 
exponents $\varsigma_{\uparrow}$ and $\varsigma_{\downarrow}$ are 
equal and given by $-1$, $-2+{1\over 2}[{\xi_0\over 
2}+{1\over {\xi_0}}]^2$, and $-{1\over 2}-\eta_0[1-{\eta_0\over 
2}]$, respectively. Here the $m\rightarrow 0$ parameter $\xi_0$ 
changes from $\xi_0=\sqrt{2}$ at $U=0$ to $\xi_0=1$ as 
$U\rightarrow\infty$ and $\eta_0=({2\over {\pi}})\tan^{-1}\left({4t\sin 
(\pi n)\over U}\right)$. 

The evaluation in Appendix C for the matrix elements of the rhs 
of expression $(43)$ refers to the thermodynamic 
limit and follows the study of the small-$\omega $ dependencies
of the one-electron Green function $G_{\sigma} (\pm k_{F\sigma},
\omega)$ and self energy $\Sigma_{\sigma} (\pm k_{F\sigma},
\omega)$. This leads to $\omega $ dependent quantities
[as $(46)$ and $(49)$ and the function $F_{\sigma}^{\alpha 
,\iota}(\omega )$ of Eq. $(51)$ below] whose $\omega\rightarrow 
0$ limits provide the expressions for these matrix elements.
Although these matrix elements vanish, it is physicaly
important to consider the associate $\omega $-dependent
functions. These are matrix-element expressions
only in the limit $\omega\rightarrow 0$, yet at small finite 
values of $\omega $ they provide revelant information 
on the electron - quasiparticle overlap at low energy
$\omega $. In addition to expression $(44)$, in Appendix C 
we find the following expression which is valid only for matrix
elements involving the excited states of form $(29)$ referring 
to the conformal-field-theory critical point

\begin{eqnarray}
\langle \{N_{ph}^{\alpha ,\iota}\},l;k=0|
c^{\dag }_{k_{F\sigma},\sigma}|0;i\rangle & = & \lim_{\omega\to 0}
F_{\sigma}^{\alpha ,\iota}(\omega ) = 0\, ,\nonumber\\
F_{\sigma}^{\alpha ,\iota}(\omega ) & = &
e^{i\chi_{\sigma }(\{N_{ph}^{\alpha ,\iota}\},l)}
\sqrt{{a^{\sigma }(\{N_{ph}^{\alpha ,\iota}\},l)\over 
a^{\sigma }_0}}\sqrt{Z_{\sigma }(\omega )}\,
\omega^{\sum_{\alpha ,\iota} N_{ph}^{\alpha ,\iota}} 
\, .
\end{eqnarray}
Here $\chi_{\sigma }(\{N_{ph}^{\alpha ,\iota}\},l)$  
and $a^{\sigma }(\{N_{ph}^{\alpha ,\iota}\},l)$ are real 
numbers and the function $Z_{\sigma }(\omega )$ was defined above. 
Notice that the function $F_{\sigma}^{\alpha ,\iota}(\omega )$
vanishes with different powers of $\omega $
for different sets of $N_{ph}^{\alpha ,\iota}$
numbers. This is because these powers reflect
directly the order of the pseudoparticle-pseudohole
generator relative to the final ground state 
of the corresponding state I. 

Although the renormalization factor $(45)$ and matrix 
elements $(51)$ vanish, Eqs. $(49)$ and $(51)$ provide relevant
information in what concerns the ratios of the different 
matrix elements which can either diverge or vanish.
Moreover, in the evaluation of some $\omega $-dependent quantities 
we can use for the matrix elements $(51)$ the function
$F_{\sigma}^{\alpha ,\iota}(\omega )$ and assume that $\omega $ 
is vanishing small, which leads to correct results.
This procedure is similar to replacing the renormalization
factor $(45)$ by the function $(49)$. While the
renormalization factor is zero because in the limit of 
vanishing excitation energy there is no overlap between 
the electron and the quasiparticle, the
function $(49)$ is associated with the small
electron - quasiparticle overlap which occurs at low excitation
energy $\omega $. 

Obviously, if we introduced in the rhs of Eq. $(43)$ zero for the 
matrix elements $(44)$ and $(51)$ we would loose all information
on the associate low-energy singular electron - quasiparticle
transformation (described by Eq. $(58)$ below). The vanishing of 
the matrix elements $(44)$ and $(51)$ just reflects the fact that 
the one-electron density of states vanishes in the 1D many-electron 
problem when the excitation energy $\omega\rightarrow 0$. 
This justifies the lack of electron - quasiparticle
overlap in the limit of zero excitation energy.
However, the diagonalization of that problem
absorbes the renormalization factor $(45)$ and
maps vanishing electronic spectral weight onto
finite quasiparticle and pseudoparticle spectral weight.
This process can only be suitable described 
if we keep either ${1\over {N_a}}$ corrections in the case
of the large finite system or small
virtual $\omega $ corrections in the case of the 
infinite system. (The analysis of Appendix C has
considered the thermodynamic limit and, therefore,
we consider in this section the case of the infinite system.)

In spite of the vanishing of the matrix elements $(44)$ and
$(51)$, following the above discussion we introduce 
Eqs. $(44)$ and $(51)$ in Eq. $(43)$ with the result

\begin{eqnarray}
c^{\dag }_{\pm k_{F\sigma},\sigma } & = & 
\lim_{\omega\to 0} \sqrt{Z_{\sigma }(\omega )}
\left[1 + \sum_{\{N_{ph}^{\alpha ,\iota}\},l}
e^{i\chi_{\sigma }(\{N_{ph}^{\alpha ,\iota}\},l)}
\sqrt{{a^{\sigma }(\{N_{ph}^{\alpha ,\iota}\},l)\over 
a^{\sigma }_0}}\omega^{\sum_{\alpha 
,\iota} N_{ph}^{\alpha ,\iota }}\prod_{\alpha=c,s}
{\hat{L}}^{\alpha\iota}_{-N_{ph}^{\alpha\iota}}(l)\right]
\tilde{c}^{\dag }_{\pm k_{F\sigma},\sigma } 
\nonumber\\
& + & \sum_{\gamma '}\langle \gamma ';k=0|c^{\dag 
}_{\pm k_{F\sigma},\sigma }|0;i\rangle {\hat{A}}_{\gamma '}
\tilde{c}^{\dag }_{\pm k_{F\sigma},\sigma } \, .
\end{eqnarray}
(Note that the expression is the same for momenta
$k=k_{F\sigma}$ and $k=-k_{F\sigma}$.) 

Let us confirm the key role played by the ``bare'' 
quasiparticle ground-state -- ground-state transition
in the low-energy physics. Since the $k=0$ higher-energy 
LWS's I and finite-energy LWS's II and non-LWS's represented 
in Eq. $(52)$ by $|\gamma ';k=0\rangle$ are irrelevant for the 
low-energy physics, we focus our attention on the
lowest-energy states of form $(29)$.

Let us look at the leading-order terms of the first term
of the rhs of Eq. $(52)$. These correspond to the 
ground-state -- ground-state transition and to the first-order 
pseudoparticle-pseudohole corrections. These corrections
are determined by the excited states $(39)-(42)$.
The use of Eqs. $(34)$ and $(39)-(42)$ allows us rewriting 
the leading-order terms as 

\begin{equation}
\lim_{\omega\to 0}\sqrt{Z_{\sigma }(\omega )}\left[1 + 
\omega^2\sum_{\alpha ,\alpha ',\iota}
C_{\alpha ,\alpha '}^{\iota }
\rho_{\alpha\iota } (\iota{2\pi\over {N_a}})
\rho_{\alpha '-\iota } (-\iota{2\pi\over {N_a}}) 
+ {\cal O}(\omega^4)\right]
\tilde{c}^{\dag }_{\pm k_{F\sigma},\sigma } \, ,
\end{equation}
where $C_{\alpha ,\alpha '}^{\iota }$ are complex constants 
such that

\begin{equation}
C_{c,c}^{1} = C_{c,c}^{-1} = e^{i\chi_{\sigma }(1,1,0,0)}
\sqrt{{a^{\sigma }(1,1,0,0)\over 
a^{\sigma }_0}} \, ,
\end{equation}

\begin{equation}
C_{s,s}^{1} = C_{s,s}^{-1} = e^{i\chi_{\sigma }(0,0,1,1)}
\sqrt{{a^{\sigma }(0,0,1,1)\over 
a^{\sigma }_0}} \, ,
\end{equation}

\begin{equation}
C_{c,s}^{1} = C_{s,c}^{-1} = 
e^{i\chi_{\sigma }(1,0,0,1)}
\sqrt{{a^{\sigma }(1,0,0,1)\over 
a^{\sigma }_0}} \, ,
\end{equation}

\begin{equation}
C_{c,s}^{-1} = C_{s,c}^{1} = 
e^{i\chi_{\sigma }(0,1,1,0)}
\sqrt{{a^{\sigma }(0,1,1,0)\over 
a^{\sigma }_0}} \, ,
\end{equation}
and ${\hat{\rho}}_{\alpha ,\iota } (k)=\sum_{\tilde{q}} 
b^{\dag}_{\tilde{q}+k,\alpha ,\iota}b_{\tilde{q},\alpha ,\iota}$ 
is a first-order pseudoparticle-pseudohole operator. 
The real constants $a^{\sigma }$ and $\chi_{\sigma }$   
in the rhs of Eqs. $(54)-(57)$ are particular cases of the
corresponding constants of the general expression $(51)$.
Note that the $l$ independence of the states $(39)-(42)$ 
allowed the omission of the index $l$ in the quantities of 
the rhs of Eqs. $(54)-(57)$ and that we used the notation 
$(28)$ for the argument of the corresponding $l$-independent
$a^{\sigma }(\{N_{ph}^{\alpha ,\iota}\})$ constants
and $\chi_{\sigma }(\{N_{ph}^{\alpha ,\iota}\})$ phases. 

The higher-order contributions to expression $(53)$ 
are associated with low-energy excited Hamiltonian 
eigenstates I orthogonal both to the initial and final
ground states and whose matrix-element amplitudes
are given by Eq. $(51)$. The corresponding functions  
$F_{\sigma}^{\alpha ,\iota}(\omega )$
vanish as $\lim_{\omega\to 0}\omega^{1+\varsigma_{\sigma}+4j\over 2}$
(with $2j$ the number of pseudoparticle-pseudohole
processes relative to the final ground state 
and $j=1,2,...$). Therefore, the leading-order term
of $(52)-(53)$ and the exponent $\varsigma_{\sigma}$ 
$(47)-(48)$ fully control the low-energy overlap between the 
$\pm k_{F\sigma}$ quasiparticles and electrons and determines the 
expressions of all $k=\pm k_{F\sigma }$ one-electron 
low-energy quantities. That leading-order term refers to
the ground-state -- ground-state transition which dominates
the electron - quasiparticle transformation $(24)$.
This transition corresponds to the ``bare'' quasiparticle 
of Eq. $(8)$. We follow the same steps as Fermi liquid 
theory and consider the low-energy non-canonical and non-complete 
transformation one derives from the full expression $(53)$ by 
only taking the corresponding leading-order term which leads to

\begin{equation}
{\tilde{c}}^{\dag }_{\pm k_{F\sigma},\sigma } =
{c^{\dag }_{\pm k_{F\sigma},\sigma }\over {\sqrt{Z_{\sigma }}}} \, .
\end{equation}
This relation refers to a singular transformation. 
Combining Eqs. $(21)-(22)$ and $(58)$ provides the low-energy 
expression for the electron in the pseudoparticle basis. The 
singular nature of the transformation $(58)$ which maps the 
vanishing-renormalization-factor electron onto the 
one-renormalization-factor quasiparticle, explains the perturbative 
character of the pseudoparticle-operator basis 
\cite{Carmelo94,Carmelo94b,Carmelo94c}.

If we replace in Eq. $(58)$ the renormalization factor
$Z_{\sigma }$ by $Z_{\sigma }(\omega )$ or omit 
$\lim_{\omega\to 0}$ from the rhs of Eqs. $(52)$ and $(53)$ 
and in both cases consider $\omega$ being very small leads to 
effective expressions which contain information on the 
low-excitation-energy electron -- quasiparticle overlap. 
Since these expressions correspond to the infinite
system, the small $\omega $ finite contributions contain
the same information as the ${1\over {N_a}}$ corrections
of the corresponding large but finite system at
$\omega =0$.

It is the perturbative character of the pseudoparticle basis
that determines the form of expansion $(53)$ which except for
the non-classical exponent in the $\sqrt{Z_{\sigma }(\omega )}
\propto \omega^{1+\varsigma_{\sigma}\over 2}$ 
factor [absorbed by the electron - quasiparticle transformation
$(58)$] includes only classical exponents, as in a Fermi liquid
\cite{Nozieres}. At low energy the BA solution performs
the singular transformation $(58)$ which absorbes the one-electron 
renormalization factor $(45)$ and maps vanishing electronic 
spectral weight onto finite quasiparticle and pseudoparticle 
spectral weight. By that process the transformation $(58)$ renormalizes 
divergent two-electron scattering vertex functions onto finite 
two-quasiparticle scattering quantities. These quantities are 
related to the finite $f$ functions \cite{Carmelo92} of form given 
by Eq. (A4) and amplitudes of scattering \cite{Carmelo92b} of the 
pseudoparticle theory. 

It was shown in Refs. \cite{Carmelo92,Carmelo92b,Carmelo94b}
that these $f$ functions and amplitudes of scattering determine 
all static and low-energy quantities of the 1D many-electron 
problem, as we discuss below and in Appendices A and D. The $f$ functions 
and amplitudes are associated with zero-momentum two-pseudoparticle
forward scattering. These scattering processes interchange
no momentum and no energy, only giving rise to two-pseudoparticle
phase shifts. The corresponding pseudoparticles control
all the low-energy physics. In the limit of vanishing energy
the pseudoparticle spectral weight leads to finite values for
the static quantities, yet it corresponds to vanishing one-electron 
spectral weight. 

To diagonalize the problem at lowest energy is equivalent to 
perform the electron - quasiparticle transformation $(58)$: 
it maps divergent irreducible (two-momenta) charge and spin 
vertices onto finite quasiparticle parameters by absorbing 
$Z_{\sigma }$. In a diagramatic picture this amounts by 
multiplying each of these vertices appearing in the diagrams
by $Z_{\sigma }$ and each one-electron Green function
(propagator) by ${1\over Z_{\sigma }}$. This procedure
is equivalent to renormalize the electron quantities onto
corresponding quasiparticle quantities, as in a Fermi
liquid. However, in the present case the renormalization
factor is zero. 

This also holds true for more involved 
four-momenta divergent two-electron vertices at the Fermi 
points. In this case the electron - quasiparticle
transformation multiplies each of these vertices
by a factor $Z_{\sigma }Z_{\sigma '}$, the factors
$Z_{\sigma }$ and $Z_{\sigma '}$ corresponding to
the pair of $\sigma $ and $\sigma '$ interacting electrons. 
The obtained finite parameters control all static 
quantities. Performimg the transformation $(58)$ is equivalent to sum 
all vertex contributions and we find that this 
transformation is unique, ie it maps the divergent Fermi-surface 
vertices on the same finite quantities independently on 
the way one chooses to approach
the low energy limit. This cannot be detected by looking only
at logarithmic divergences of some diagrams \cite{Solyom,Metzner}. 
Such non-universal contributions either cancel or
are renormalized to zero by the electron - quasiparticle
transformation. We have extracted all our results from the exact 
BA solution which takes into account all relevant contributions.
We can choose the energy variables in such a way that
there is only one $\omega $ dependence. We find that the 
relevant vertex function divergences are controlled by the 
electron - quasiparticle overlap, the vertices reading  

\begin{equation}
\Gamma_{\sigma\sigma '}^{\iota }(k_{F\sigma },\iota 
k_{F\sigma '};\omega) = {1\over 
{Z_{\sigma}(\omega)Z_{\sigma '}(\omega)}}
\{\sum_{\iota '=\pm 1}(\iota ')^{{1-\iota\over 2}}
[v_{\rho }^{\iota '} + (\delta_{\sigma ,\sigma '}
- \delta_{\sigma ,-\sigma '})v_{\sigma_z}^{\iota '}]
- \delta_{\sigma ,\sigma '}v_{F,\sigma }\}
\, ,
\end{equation}
where the expressions for the charge $v_{\rho}^{\iota }$ and 
spin $v_{\sigma_z}^{\iota }$ velocities
are given in Appendix D. The divergent character of the function 
$(59)$ follows exclusively from the ${1\over Z_{\sigma}(\omega)Z_{\sigma 
'}(\omega)}$ factor, with $Z_{\sigma}(\omega)$ given by $(49)$.
The transformation $(58)$ maps the divergent vertices onto the
$\omega $-independent finite quantity $Z_{\sigma}(\omega) 
Z_{\sigma '}(\omega)\Gamma_{\sigma\sigma '}^{\iota }(k_{F\sigma },
\iota k_{F\sigma '};\omega )$. The low-energy physics is determined by
the following $v_{F,\sigma }$-independent 
Fermi-surface two-quasiparticle parameters

\begin{equation}
L^{\iota }_{\sigma ,\sigma'} = lim_{\omega\to 0}
\left[\delta_{\sigma ,\sigma '}v_{F,\sigma }+
Z_{\sigma}(\omega) Z_{\sigma '}(\omega)\Gamma_{\sigma\sigma 
'}^{\iota }(k_{F\sigma },\iota k_{F\sigma '};\omega )\right] \, .
\end{equation}
From the point of view of the electron - quasiparticle
transformation the divergent vertices $(59)$ originate
the finite quasiparticle parameters $(60)$ which define the 
above charge and spin velocities. These are given by the 
following simple combinations of the parameters $(60)$

\begin{eqnarray}
v_{\rho}^{\iota} = {1\over 4}\sum_{\iota '=\pm 1}(\iota 
')^{{1-\iota\over 2}}\left[L_{\sigma ,\sigma}^{\iota '} + 
L_{\sigma ,-\sigma}^{\iota '}\right] 
\, , \nonumber\\
v_{\sigma_z}^{\iota} = {1\over 4}\sum_{\iota '=\pm 1}(\iota 
')^{{1-\iota\over 2}}\left[L_{\sigma ,\sigma}^{\iota '} - 
L_{\sigma ,-\sigma}^{\iota '}\right] \, .
\end{eqnarray}
As shown in Appendix D, the parameters $L_{\sigma ,\sigma '}^{\iota}$ 
can be expressed in terms of the pseudoparticle group velocities
(A6) and Landau parameters (A8) as follows

\begin{eqnarray}
L_{\sigma ,\sigma}^{\pm 1} & = & 2\left[{(v_s + F^0_{ss})\over L^0}
\pm (v_c + F^1_{cc}) - {L_{\sigma ,-\sigma}^{\pm 1}\over 2}
\right] \, , \nonumber\\
L_{\sigma ,-\sigma}^{\pm 1} & = & -4\left[{(v_c + F^0_{cc} + 
F^0_{cs})\over L^0}\pm (v_s + F^1_{ss} - F^1_{cs})\right] 
\, ,
\end{eqnarray}
where $L^0=(v_c+F^0_{cc})(v_s+F^0_{ss})-(F^0_{cs})^2$.
Combining equations $(61)$ and $(62)$ we find the expressions
of the Table for the charge and spin velocities. These velocities 
were already known through the BA solution and determine the 
expressions for all static quantities \cite{Carmelo94c}. 
Equations $(62)$ clarify their origin which is the singular
electron - quasiparticle transformation $(58)$. It renders a 
non-perturbative electronic problem into a perturbative 
pseudoparticle problem. In Appendix D
we show how the finite two-pseudoparticle forward-scattering 
$f$ functions and amplitudes which determine the static quantities are 
directly related to the two-quasiparticle finite parameters
$(60)$ through the velocities $(61)$. This study confirms that it 
is the singular electron - quasiparticle transformation $(58)$ 
which justifies the {\it finite character} of the $f_{\alpha\alpha 
'}(q,q')$ functions (A4) and the associate perturbative origin of 
the pseudoparticle Hamiltonian $(6)-(7)$ \cite{Carmelo94}. 

In order to further confirm that the electron - quasiparticle
transformation $(58)$ and associate electron - quasiparticle
overlap function $(49)$ control the whole low-energy
physics we close this section by considering the one-electron
spectral function. The spectral function 
was studied numerically and for $U\rightarrow\infty$ in Refs. 
\cite{Muramatsu} and \cite{Shiba}, respectively. 
The leading-order term of the real-part expression
for the $\sigma $ Green function at $k=\pm k_{F\sigma}$
and small excitation energy $\omega $ (C10)-(C11) is given 
by, $\hbox{Re}G_{\sigma} (\pm k_{F\sigma},\omega)=a^{\sigma}_0
\omega^{\varsigma_{\sigma}}$. From Kramers-Kronig relations 
we find $\hbox{Im}G_{\sigma} (\pm k_{F\sigma},\omega)=
-i\pi a^{\sigma}_0 (1 + \varsigma_{\sigma})
\omega^{\varsigma_{\sigma }}$ for the corresponding 
imaginary part. Based on these results we arrive to the
following expression for the low-energy spectral
function at $k=\pm k_{F\sigma}$

\begin{equation}
A_{\sigma}(\pm k_{F\sigma},\omega) =
2\pi a^{\sigma }_0 (1 + \varsigma_{\sigma})
\omega^{\varsigma_{\sigma}} = 2\pi {\partial 
Z_{\sigma}(\omega)\over\partial\omega} \, . 
\end{equation}
This result is a generalization of the $U\rightarrow\infty$
expression of Ref. \cite{Shiba}. It is valid for all
parameter space where both the velocities $v_c$ and $v_s$ (A6) 
are finite. (This excludes half filling $n=1$, maximum spin density 
$m=n$, and $U=\infty$ when $m\neq 0$.) The use of
Kramers-Kronig relations also restricts the validity of
expression $(63)$ to the energy $\omega $ continuum limit.
On the other hand, we can show that $(63)$ is consistent with 
the general expression

\begin{eqnarray}
A_{\sigma} (\pm k_{F\sigma},\omega) 
& = & \sum_{\{N_{ph}^{\alpha ,\iota}\},l}
|\langle \{N_{ph}^{\alpha ,\iota}\},l;k=0|
c^{\dag }_{k_{F\sigma},\sigma}|0;i\rangle |^2 
2\pi\delta (\omega - \omega (\{N_{ph}^{\alpha 
,\iota}\}))\nonumber \\ 
& + & \sum_{\gamma '}|\langle\gamma ';k=0|c^{\dag }_{k_{F\sigma},\sigma}
|0;i\rangle |^2 2\pi\delta (\omega - \omega_{\gamma '}) \, ,
\end{eqnarray}
whose summations refer to the same states as the summations
of expressions $(43)$ and $(52)$.
The restriction of the validity of expression $(63)$ 
to the energy continuum limit requires the 
consistency to hold true only for the spectral weight 
of $(64)$ associated with the quasiparticle ground-state -- 
ground-state transition. This corresponds to the first 
$\delta $ peak of the rhs of Eq. $(64)$. Combining equations 
$(44)$ and $(64)$ and considering that in the present
limit of vanishing $\omega $ replacing the renormalization
factor $(45)$ by the electron - quasiparticle overlap
function $(49)$ leads to the correct result (as we
confirm below) we arrive to

\begin{equation}
A_{\sigma}(\pm k_{F\sigma},\omega) =
a^{\sigma }_0\omega^{1+\varsigma_{\sigma}} 
2\pi\delta (\omega - \omega^0_{\sigma }) 
= Z_{\sigma}(\omega ) 2\pi\delta (\omega - 
\omega^0_{\sigma }) \, .
\end{equation}
Let us then show that the Kramers-Kronig continuum expression 
$(63)$ is an approximation consistent with the Dirac-delta function 
representation $(65)$. This consistency just requires that 
in the continuum energy domain from $\omega =0$ to the
ground-state -- ground-state transition energy
$\omega =\omega^0_{\sigma }$ (see Eq. $(35)$) the functions $(63)$
and $(65)$ contain the same amount of spectral weight.
We find that both the $A_{\sigma}(\pm k_{F\sigma},\omega)$
representations $(63)$ and $(65)$ lead to

\begin{equation}
\int_{0}^{\omega^0_{\sigma }}A_{\sigma}(\pm k_{F\sigma},\omega)
=2\pi a^{\sigma }_0 [\omega^0_{\sigma }]^{\varsigma_{\sigma }+1}
\, , 
\end{equation}
which confirms they contain the same spectral weight.
The representation $(63)$ reveals that the spectral
function diverges at $\pm k_{F\sigma}$ and small $\omega$
as a Luttinger-liquid power law. However, both the small-$\omega $
density of states and the integral $(66)$ vanish
in the limit of vanishing excitation energy.

Using the method of Ref. \cite{Carmelo93} we have also
studied the spectral function $A_{\sigma}(k,\omega)$ for all
values of $k$ and vanishing positive $\omega $.
We find that $A_{\sigma}(k,\omega)$ [and 
the Green function $Re G_{\sigma} (k,\omega)$]
vanishes when $\omega\rightarrow 0$ for all momentum values 
{\it except} at the non-interacting Fermi-points 
$k=\pm k_{F\sigma}$ where it diverges as the
power law $(63)$. This divergence is fully controlled
by the quasiparticle ground-state - ground-state
transition. The transitions to the excited states
$(29)$ give only vanishing contributions to the
spectral function. This further confirms the dominant
role of the bare quasiparticle ground-state - ground-state 
transition and of the associate electron - quasiparticle
transformation $(58)$ which control the low-energy
physics.

It follows from the above behavior of the spectral function 
at small $\omega $ that for $\omega\rightarrow 0$ 
the density of states, 

\begin{equation}
D_{\sigma} (\omega)=\sum_{k}A_{\sigma}(k,\omega) \, ,
\end{equation}
results, exclusively, from contributions of
the peaks centered at $k=\pm k_{F\sigma}$ 
and is such that $D_{\sigma} (\omega)\propto 
\omega A_{\sigma}(\pm k_{F\sigma},\omega)$ \cite{Carmelo95a}.
On the one hand, it is known from the zero-magnetic
field studies of Refs. \cite{Shiba,Schulz} that
the density of states goes at small $\omega $ as

\begin{equation}
D_{\sigma} (\omega)\propto\omega^{\nu_{\sigma}} \, ,
\end{equation}
where $\nu_{\sigma}$ is the exponent of 
the equal-time momentum distribution expression, 

\begin{equation}
N_{\sigma}(k)\propto |k\mp k_{F\sigma}|^{\nu_{\sigma }} \, ,
\end{equation}
\cite{Frahm91,Ogata}. (The exponent $\nu_{\sigma }$ is defined by Eq. 
$(5.10)$ of Ref. \cite{Frahm91} for the particular case of the 
$\sigma $ Green function.) On the other hand, we find that the 
exponents $(47)-(48)$ and $\nu_{\sigma}$ are such that 

\begin{equation}
\varsigma_{\sigma}=\nu_{\sigma }-1 \, , 
\end{equation}
in agreement with the above analysis. However, this simple relation 
does not imply that the equal-time expressions \cite{Frahm91,Ogata}
provide full information on the small-energy instabilities. 
For instance, in addition to the momentum values
$k=\pm k_{F\sigma}$ and in contrast to the spectral function,  
$N_{\sigma}(k)$ shows singularities at
$k=\pm [k_{F\sigma}+2k_{F-\sigma}]$ \cite{Ogata}.
Therefore, only the direct low-energy study reveals all the 
true instabilities of the quantum liquid.

Note that in some Luttinger liquids the momentum
distribution is also given by $N(k)\propto |k\mp k_F|^{\nu }$
but with $\nu >1$ \cite{Solyom,Medem,Voit}. We find that in these 
systems the spectral function $A(\pm k_F,\omega)
\propto\omega^{\nu -1}$ does not diverge.

%%%%%%%%%%%%%%%%%%%%%%%%%%%%%%%%%%%%%%%%%%%%%%%%%%%%%%%%%%%%%%%%
\section{CONCLUDING REMARKS}

One of the goals of this paper was, in spite
of the differences between the Luttinger-liquid
Hubbard chain and 3D Fermi liquids, 
detecting common features in these two limiting
problems which we expect to be present in electronic
quantum liquids in spatial dimensions
$1<$D$<3$. As in 3D Fermi liquids, we find that
there are Fermi-surface quasiparticles in
the Hubbard chain which connect ground states 
differing in the number of electrons by one and
whose low-energy overlap with electrons
determines the $\omega\rightarrow 0$
divergences. In spite of the vanishing
electron density of states and renormalization
factor, the spectral function vanishes at all momenta 
values {\it except} at the Fermi surface where it 
diverges (as a Luttinger-liquid power law).

While low-energy excitations are
described by $c$ and $s$ pseudoparticle-pseudohole
excitations which determine the $c$ and $s$
separation \cite{Carmelo94c}, the quasiparticles 
describe ground-state -- ground-state transitions 
and recombine $c$ and $s$ (charge and spin in 
the zero-magnetization limit), being labelled by the
spin projection $\sigma $. They are constituted
by one topological momenton and one or two
pseudoparticles which cannot be separated and
are confined inside the quasiparticle.
Moreover, there is a close relation between the quasiparticle
contents and the Hamiltonian symmetry in the different
sectors of parameter space. This can be shown 
if we consider pseudoholes instead of pseudoparticles
\cite{Carmelo95a} and we extend the present quasiparticle
study to the whole parameter space of the Hubbard chain.

Importantly, we have written the low-energy electron 
at the Fermi surface in the pseudoparticle basis.
The vanishing of the electron renormalization
factor implies a singular character for the low-energy
electron -- quasiparticle and electron --
pseudoparticle transformations. This singular process extracts 
from vanishing electron spectral weight quasiparticles of 
spectral-weight factor one.
The BA diagonalization of the 1D many-electron problem
is at lowest excitation energy equivalent to perform
such singular electron -- quasiparticle transformation.
This absorves the vanishing one-electron renormalization
factor giving rise to the finite two-pseudoparticle forward-scattering 
$f$ functions and amplitudes which control the expressions for 
all static quantities \cite{Carmelo92,Carmelo92b,Carmelo94}.
It is this transformation which justifies the 
perturbative character of the many-electron Hamiltonian in the 
pseudoparticle basis \cite{Carmelo94}. 

From the existence of Fermi-surface 
quasiparticles both in the 1D and 3D limits, our results 
suggest their existence for quantum liquids in dimensions 
1$<$D$<$3. However, the effect of increasing dimensionality 
on the electron -- quasiparticle overlap remains an
unsolved problem. The present 1D results do not provide
information on whether that overlap can vanish for D$>$1
or whether it always becomes finite as soon as we leave 1D.

%%%%%%%%%%%%%%%%%%%%%%%%%%%%%%%%%%%%%%%%%%%%%%%%%%%%%%%%%%%%%%%%%%%
\nonum
\section{ACKNOWLEDGMENTS}

We thank N. M. R. Peres for many fruitful discussions and for 
reproducing and checking some of our calculations. We are 
grateful to F. Guinea and K. Maki for illuminating discussions. 
This research was supported in part by the Institute for Scientific 
Interchange Foundation under the EU Contract No. ERBCHRX - CT920020 
and by the National Science Foundation under the Grant No. 
PHY89-04035.

%%%%%%%%%%%%%%%%%%%%%%%%%%%%%%%%%%%%%%%%%%%%%%%%%%%%%%%%%%%%%%%%%%%
\vfill
\eject
\appendix{SOME USEFUL QUANTITIES OF THE PSEUDOPARTICLE
REPRESENTATION}

In this Appendix we present some quantities of the pseudoparticle
picture which are useful for the present study. We start
by defining the pseudo-Fermi points and limits of
the pseudo-Brillouin zones. When $N_{\alpha }$ (see Eq. $(4)$)
is odd (even) and the numbers $I_j^{\alpha }$ of Eq. $(3)$ are 
integers (half integers) the pseudo-Fermi 
points are symmetric and given by \cite{Carmelo94,Carmelo94c}

\begin{equation}
q_{F\alpha }^{(+)}=-q_{F\alpha }^{(-)} =
{\pi\over {N_a}}[N_{\alpha}-1] \, .
\end{equation}
On the other hand, when $N_{\alpha }$ is odd (even) and
$I_j^{\alpha }$ are half integers (integers)
we have that

\begin{equation}
q_{F\alpha }^{(+)} = {\pi\over {N_a}}N_{\alpha }
\, , \hspace{1cm} 
-q_{F\alpha }^{(-)} ={\pi\over {N_a}}[N_{\alpha }-2] \, ,
\end{equation}
or 

\begin{equation}
q_{F\alpha }^{(+)} = {\pi\over {N_a}}[N_{\alpha }-2] 
\, , \hspace{1cm} 
-q_{F\alpha }^{(-)} = {\pi\over {N_a}}N_{\alpha } \, . 
\end{equation}
Similar expressions are obtained for the pseudo-Brioullin 
zones limits $q_{\alpha }^{(\pm)}$ if we replace
in Eqs. (A1)-(A3) $N_{\alpha }$ by the numbers 
$N_{\alpha }^*$ of Eq. $(4)$. 

The $f$ functions were studied in Ref. \cite{Carmelo92}
and read

\begin{eqnarray}
f_{\alpha\alpha'}(q,q') & = & 2\pi v_{\alpha}(q) 
\Phi_{\alpha\alpha'}(q,q')  
+ 2\pi v_{\alpha'}(q') \Phi_{\alpha'\alpha}(q',q) \nonumber \\
& + & \sum_{j=\pm 1} \sum_{\alpha'' =c,s}
2\pi v_{\alpha''} \Phi_{\alpha''\alpha}(jq_{F\alpha''},q)
\Phi_{\alpha''\alpha'}(jq_{F\alpha''},q') \, ,
\end{eqnarray}
where the pseudoparticle group velocities are given by 

\begin{equation}
v_{\alpha}(q) = {d\epsilon_{\alpha}(q) \over {dq}} \, , 
\end{equation}
and 

\begin{equation}
v_{\alpha }=\pm 
v_{\alpha }(q_{F\alpha}^{(\pm)}) \, ,
\end{equation}
are the pseudo-Fermi points group velocities. In expression
(A4) $\Phi_{\alpha\alpha '}(q,q')$ mesures the phase shift
of the $\alpha '$ pseudoparticle of pseudomomentum $q'$
due to the forward-scattering collision with the
$\alpha $ pseudoparticle of pseudomomentum $q$.
These phase shifts determine the pseudoparticle
interactions and are defined in Ref. \cite{Carmelo92}.
They control the low-energy physics. For instance, the 
related parameters

\begin{equation}
\xi_{\alpha\alpha '}^j = \delta_{\alpha\alpha '}+ 
\Phi_{\alpha\alpha '}(q_{F\alpha}^{(+)},q_{F\alpha '}^{(+)})+
(-1)^j\Phi_{\alpha\alpha '}(q_{F\alpha}^{(+)},q_{F\alpha '}^{(-)})
\, , \hspace{2cm} j=0,1 \, ,
\end{equation}
play a determining role at the critical point. 
($\xi_{\alpha\alpha '}^1$ are the entries of the transpose
of the dressed-charge matrix \cite{Frahm}.)  
The values at the pseudo-Fermi points of the $f$
functions (A4) include the parameters (A7) 
and define the Landau parameters, 

\begin{equation}
F_{\alpha\alpha'}^j = 
{1\over {2\pi}}\sum_{\iota =\pm 1}(\iota )^j
f_{\alpha\alpha'}(q_{F\alpha}^{(\pm)},\iota 
q_{F\alpha '}^{(\pm)}) \, , \hspace{1cm} j=0,1 \, .
\end{equation}
These are also studied in Ref. \cite{Carmelo92}.
The parameters $\delta_{\alpha ,\alpha'}v_{\alpha }+
F_{\alpha\alpha'}^j$ appear in the expressions of the low-energy 
quantities.

We close this Appendix by introducing pseudoparticle-pseudohole
operators which will appear in Sec. IV.
Although the expressions in the pseudoparticle basis
of one-electron operators remains an unsolved
problem, in Ref. \cite{Carmelo94c} the electronic fluctuation
operators 

\begin{equation}
{\hat{\rho}}_{\sigma }(k)=
\sum_{k'}c^{\dag }_{k'+k\sigma}c_{k'\sigma} \, ,
\end{equation}
were expressed in terms of the pseudoparticle fluctuation operators

\begin{equation}
{\hat{\rho}}_{\alpha }(k)=\sum_{q}b^{\dag }_{q+k\alpha}
b_{q\alpha} \, .
\end{equation}
This study has revealed that $\iota =sgn (k)1=\pm 1$ electronic 
operators are made out of $\iota =sgn (q)1=\pm 1$ pseudoparticle 
operators only, $\iota $ defining the right ($\iota=1$) and left 
($\iota=-1$) movers.

Often it is convenient measuring the electronic momentum $k$ and 
pseudomomentum $q$ from the $U=0$ Fermi points 
$k_{F\sigma}^{(\pm)}=\pm \pi n_{\sigma}$ and  pseudo-Fermi 
points $q_{F\alpha}^{(\pm)}$, respectively. This adds the index 
$\iota$ to the electronic and pseudoparticle operators.
The new momentum $\tilde{k}$ and pseudomomentum $\tilde{q}$ are 
such that 

\begin{equation}
\tilde{k} =k-k_{F\sigma}^{(\pm)} \, , \hspace{2cm}
\tilde{q}=q-q_{F\alpha}^{(\pm)} \, , 
\end{equation}
respectively, for $\iota=\pm 1$.
For instance, 

\begin{equation}
{\hat{\rho}}_{\sigma ,\iota }(k)=\sum_{\tilde{k}}
c^{\dag }_{\tilde{k}+k\sigma\iota}c_{\tilde{k}\sigma\iota} \, ,
\hspace{2cm}
{\hat{\rho}}_{\alpha ,\iota }(k) =
\sum_{\tilde{q}}b^{\dag }_{\tilde{q}+k\alpha\iota}
b_{\tilde{q}\alpha\iota} \, . 
\end{equation}

%%%%%%%%%%%%%%%%%%%%%%%%%%%%%%%%%%%%%%%%%%%%%%%%%%%%%%%%%%%%%%%%%%%
\vfill
\eject
\appendix{THE TOPOLOGICAL-MOMENTON GENERATOR}

In this Appendix we evaluate the expression for the
topological-momenton generator
$(17)-(19)$. In order to derive the expression for $U_c^{+1}$
we consider the Fourier transform of the pseudoparticle
operator $b^{\dag}_{q,c}$ which reads

\begin{equation}
\beta^{\dag}_{x,c} = \frac{1}{\sqrt{N_a}} 
\sum_{q_c^{(-)}}^{q_c^{(+)}}
e^{-i q x} b^{\dag}_{q,c} \, .
\end{equation}
From Eq. $(15)$ we arrive to

\begin{equation}
U_c^{+1} \beta^{\dag}_{x,c} U_c^{-1} = \frac{1}{\sqrt{N_a}} 
\sum_{q_c^{(-)}}^{q_c^{(+)}}
e^{-i q x}  b^{\dag}_{q-\frac{\pi}{N_a},c} \, .
\end{equation} 
By performing a $\frac{\pi}{N_a}$ pseudomomentum
translation we find

\begin{equation}
U_c^{+1} \beta^{\dag}_{x,c} U_c^{-1} = 
e^{i\frac{\pi}{N_a} x}\beta^{\dag}_{x,c} \, , 
\end{equation} 
and it follows that 

\begin{eqnarray}
U_{c}^{\pm 1} = \exp\left\{\pm i\frac{\pi}{N_a}
\sum_{y}y\beta^{\dag}_{y,c}\beta_{y,c} \, ,
\right\}.
\end{eqnarray}
By inverse-Fourier transforming expression (B4) we
find expression $(17)-(19)$ for this unitary operator, 
which can be shown to also hold true for $U_{s}^{\pm 1}$. 

%%%%%%%%%%%%%%%%%%%%%%%%%%%%%%%%%%%%%%%%%%%%%%%%%%%%%%%%%%%%%%%%%%%
\vfill
\eject
\appendix{ONE-ELECTRON MATRIX ELEMENTS}

In this Appendix we derive the expressions for the
matrix elements $(44)$ and $(51)$.

At energy scales smaller than the gaps for the LWS's II
and non-LWS's referred in this paper and in Refs. 
\cite{Carmelo94,Carmelo94b,Carmelo94c}
the expression of the $\sigma $ one-electron Green 
function $G_{\sigma} (k_{F\sigma},\omega)$ is fully
defined in the two Hilbert sub spaces spanned by
the final ground state $|0;f\rangle $ and associate
$k=0$ excited states $|\{N_{ph}^{\alpha ,\iota}\},l;k=0\rangle$
of form $(29)$ belonging the $N_{\sigma }+1$ sector and by a corresponding
set of states belonging the $N_{\sigma }-1$ sector, respectively.
Since $|0;f\rangle $ corresponds to zero values for
all four numbers $(28)$ in this Appendix we use the notation 
$|0;f\rangle\equiv|\{N_{ph}^{\alpha ,\iota}=0\},l;k=0\rangle$. 
This allows a more compact notation for the state summations. 
The use of a Lehmann representation leads to

\begin{equation}
G_{\sigma} (k_{F\sigma},\omega) =
G_{\sigma}^{(N_{\sigma }+1)} (k_{F\sigma},\omega) 
+ G_{\sigma}^{(N_{\sigma }-1)} (k_{F\sigma},\omega) \, ,
\end{equation}
where

\begin{equation}
G_{\sigma}^{(N_{\sigma }+1)} (k_{F\sigma},\omega) =
\sum_{\{N_{ph}^{\alpha ,\iota}\},l}
{|\langle \{N_{ph}^{\alpha ,\iota}\},l;k=0|
c^{\dag }_{k_{F\sigma},\sigma}|0;i\rangle |^2\over 
{\omega - \omega (\{N_{ph}^{\alpha ,\iota}\})
+ i\xi}} \, ,
\end{equation}
has divergences for $\omega >0$ and
$G_{\sigma}^{(N_{\sigma }-1)} (k_{F\sigma},\omega)$
has divergences for $\omega <0$. We emphasize that
in the $\{N_{ph}^{\alpha ,\iota}\}$ summation 
of the rhs of Eq. (C2), $N_{ph}^{\alpha ,\iota}=0$ 
for all four numbers refers to the final ground state,
as we mentioned above. Below we consider positive but vanishing values
of $\omega $ and, therefore, we need only
to consider the function (C2).
We note that at the conformal-field critical
point \cite{Frahm,Frahm91} the states which 
contribute to (C2) are such that the ratio $N_{ph}^{\alpha 
,\iota}/N_a$ vanishes in the thermodynamic limit, 
$N_a\rightarrow 0$ \cite{Carmelo94b}.
Therefore, in that limit the positive excitation
energies $\omega (\{N_{ph}^{\alpha ,\iota}\})$ of Eq. (C2),
which are of the form $(38)$, are vanishing small. Replacing 
the full Green function by (C2) (by considering positive values of
$\omega $ only) we find

\begin{equation}
\lim_{N_a\to\infty}\hbox{Re}G_{\sigma} (k_{F\sigma},\omega) =
\sum_{\{N_{ph}^{\alpha ,\iota}\}}\left[
{\sum_{l} |\langle \{N_{ph}^{\alpha ,\iota}\},l;k=0|
c^{\dag }_{k_{F\sigma},\sigma}|0;i\rangle |^2 \over {\omega }}
\right] \, .
\end{equation}

We emphasize that considering the limit (C3) implies 
that all the corresponding expressions for the $\omega $ dependent  
quantities we obtain in the following are only valid in the limit 
of vanishing positive energy $\omega $. Although many of these
quantities are zero in that limit, their $\omega $
dependence has physical meaning because different
quantities vanish as $\omega\rightarrow 0$ in 
different ways, as we discuss in Sec. IV. Therefore, our 
results allow the classification of the relative importance of the
different quantities.
			  
In order to solve the present problem we have to combine 
a suitable generator pseudoparticle analysis \cite{Carmelo94b}
with conformal-field theory \cite{Frahm,Frahm91}.
Let us derive an alternative expression for the Green function 
(C3). Comparison of both expressions leads to relevant information. 
This confirms the importance of the pseudoparticle operator
basis \cite{Carmelo94,Carmelo94b,Carmelo94c} which
allows an operator description of the conformal-field
results for BA solvable many-electron problems 
\cite{Frahm,Frahm91}. 

The asymptotic expression of the Green function in $x$ and $t$ 
space is given by the summation of many terms of form $(3.13)$ 
of Ref. \cite{Frahm} with dimensions of the fields suitable 
to that function. For small energy the Green function in $k$ and 
$\omega $ space is obtained by the summation of the Fourier 
transforms of these terms, which are of the form given by Eq. 
$(5.2)$ of Ref. \cite{Frahm91}. However, the results of 
Refs. \cite{Frahm,Frahm91} do not provide the expression 
at $k=k_{F\sigma }$ and small positive $\omega $.
In this case the above summation is equivalent to a 
summation in the final ground state and excited states of form 
$(29)$ obeying to Eqs. $(31)$, $(32)$, and $(34)$ which correspond 
to different values for the dimensions of the fields. 

We emphasize that expression $(5.7)$ of Ref. \cite{Frahm91}
is not valid in our case. Let us use the notation
$k_0=k_{F\sigma }$ (as in Eqs. $(5.6)$ and $(5.7)$ of Ref. 
\cite{Frahm91}). While we consider $(k-k_0)=(k-k_{F\sigma })=0$ 
expression $(5.7)$ of Ref. \cite{Frahm91} is only valid when
$(k-k_0)=(k-k_{F\sigma })$ is small but finite.
We have solved the following general integral 

\begin{equation}
\tilde{g}(k_0,\omega ) = \int_{0}^{\infty}dt 
e^{i\omega t}F(t) \, ,  
\end{equation}
where

\begin{equation}
F(t) = \int_{-\infty}^{\infty}dx\prod_{\alpha ,\iota}
{1\over {(x+\iota v_{\alpha }t)^{2\Delta_{\alpha 
}^{\iota}}}} \, ,  
\end{equation}
with the result

\begin{equation}
\tilde{g}(k_0,\omega )\propto
\omega^{[\sum_{\alpha ,\iota} 
2\Delta_{\alpha }^{\iota}-2]} \, .  
\end{equation}
Comparing our expression (C6) with expression $(5.7)$ of Ref.
\cite{Frahm91} we confirm these expressions
are different. 

In the present case of the final ground state and excited
states of form $(29)$ obeying Eqs. $(31)$, $(32)$, and $(34)$ 
we find that the dimensions of the fields are such that

\begin{equation}
\sum_{\alpha ,\iota} 2\Delta_{\alpha }^{\iota}=  
2+\varsigma_{\sigma}+2\sum_{\alpha ,\iota}
N_{ph}^{\alpha ,\iota} \, ,
\end{equation}
with $\varsigma_{\sigma}$ being the exponents $(47)$ and
$(48)$. Therefore, equation (C6) can be rewritten as

\begin{equation}
\tilde{g}(k_0,\omega )\propto
\omega^{\varsigma_{\sigma}+2\sum_{\alpha ,\iota}
N_{ph}^{\alpha ,\iota}} \, .  
\end{equation}
Summing the terms of form (C8) corresponding to
different states leads to an alternative expression
for the function (C3) with the result

\begin{equation}
\lim_{N_a\to\infty} \hbox{Re}G_{\sigma} (k_{F\sigma},\omega) =
\sum_{\{N_{ph}^{\alpha ,\iota}\}}\left[
{a^{\sigma }(\{N_{ph}^{\alpha ,\iota}\}) 
\omega^{\varsigma_{\sigma } + 1
+ 2\sum_{\alpha ,\iota} N_{ph}^{\alpha ,\iota} }
\over {\omega }}
\right] \, ,
\end{equation}
or from Eq. $(34)$,

\begin{equation}
\lim_{N_a\to\infty} \hbox{Re}G_{\sigma} (k_{F\sigma},\omega) =
\sum_{j=0,1,2,...}\left[
{a^{\sigma }_j \omega^{\varsigma_{\sigma } + 1 + 4j}
\over {\omega }}\right] \, ,
\end{equation}
where $a^{\sigma }(\{N_{ph}^{\alpha ,\iota}\})$ and
$a^{\sigma }_j$ are complex constants. From equation (C10) 
we find

\begin{equation}
\hbox{Re}\Sigma_{\sigma} ( k_{F\sigma},\omega ) =
\omega - {1\over {\hbox{Re} G_{\sigma} (k_{F\sigma },\omega )}}
= \omega [1-{\omega^{-1-\varsigma_{\sigma}}\over
{a^{\sigma }_0+\sum_{j=1}^{\infty}a^{\sigma }_j\omega^{4j}}}] 
\, .
\end{equation}

While the function $Re G_{\sigma} (k_{F\sigma},\omega)$ 
(C9)-(C10) diverges as $\omega\rightarrow 0$, following the form
of the self energy (C11) the one-electron renormalization factor
$(45)$ vanishes and there is no overlap between the
quasiparticle and the electron, in contrast to a Fermi liquid.
(In equation (C11) $\varsigma_{\sigma}\rightarrow -1$ and
$a^{\sigma }_0\rightarrow 1$ when $U\rightarrow 0$.)

Comparision of the terms of expressions (C3) and (C9)
with the same $\{N_{ph}^{\alpha ,\iota}\}$ values, which
refer to contributions from the same set of 
$N_{\{N_{ph}^{\alpha ,\iota}\}}$
Hamiltonian eigenstates $|\{N_{ph}^{\alpha 
,\iota}\},l;k=0\rangle$ and refer to the limit
$\omega\rightarrow 0$, leads to

\begin{equation}
\sum_{l} |\langle \{N_{ph}^{\alpha ,\iota}\},l;k=0|
c^{\dag }_{k_{F\sigma},\sigma}|0;i\rangle |^2 =
\lim_{\omega\to 0} a^{\sigma }(\{N_{ph}^{\alpha ,\iota}\})\, 
\omega^{\varsigma_{\sigma } + 1
+ 2\sum_{\alpha ,\iota} N_{ph}^{\alpha ,\iota}} = 0\, .
\end{equation}
Note that the functions of the rhs of Eq. (C12) corresponding to
different matrix elements go to zero with different 
exponents.

On the other hand, as for the corresponding excitation
energies $(38)$, the dependence of functions associated
with the amplitudes
$|\langle \{N_{ph}^{\alpha ,\iota}\},l;k=0|
c^{\dag }_{k_{F\sigma},\sigma}|0;i\rangle |$ on
the vanishing energy $\omega $ is $l$ independent.
Therefore, we find

\begin{equation}
|\langle \{N_{ph}^{\alpha ,\iota}\},l;k=0|
c^{\dag }_{k_{F\sigma},\sigma}|0;i\rangle |^2 =
\lim_{\omega\to 0}
a^{\sigma }(\{N_{ph}^{\alpha ,\iota}\},l)\,
\omega^{\varsigma_{\sigma } + 1
+ 2\sum_{\alpha ,\iota} N_{ph}^{\alpha ,\iota}} = 0\, ,
\end{equation}
where the constants $a^{\sigma }(\{N_{ph}^{\alpha 
,\iota}\},l)$ are $l$ dependent and obey the
normalization condition

\begin{equation}
a^{\sigma }(\{N_{ph}^{\alpha ,\iota}\}) =
\sum_{l} a^{\sigma }(\{N_{ph}^{\alpha ,\iota}\},l) \, .
\end{equation}
It follows that the matrix elements of Eq. (C12) have
the form given in Eq. $(51)$.

Moreover, following our notation for the final ground
state when the four $N_{ph}^{\alpha ,\iota}$ vanish
Eq. (C13) leads to

\begin{equation}
|\langle 0;f|c^{\dag }_{k_{F\sigma},\sigma}|0;i\rangle |^2 =
\lim_{\omega\to 0}
a^{\sigma }_0\,\omega^{\varsigma_{\sigma } + 1} = 
\lim_{\omega\to 0} Z_{\sigma}(\omega) = Z_{\sigma} = 0 \, ,
\end{equation}
where $a^{\sigma }_0=a^{\sigma }(\{N_{ph}^{\alpha ,\iota}=0\},l)$
is a positive real constant and $Z_{\sigma}(\omega)$ is the
function $(49)$. Following equation (C11) the function
$Z_{\sigma}(\omega)$ is given by the leading-order term
of expression $(46)$. Since $a^{\sigma }_0$ is real
and positive expression $(44)$ follows from Eq. (C15).

%%%%%%%%%%%%%%%%%%%%%%%%%%%%%%%%%%%%%%%%%%%%%%%%%%%%%%%%%%%%%%%%%%%
\vfill
\eject
\appendix{DIVERGENT TWO-ELECTRON AND FINITE TWO-PSEUDOPARTICLE
QUANTITIES}

In this Appendix we confirm that the finite two-quasiparticle
functions $(60)$ of form $(62)$ which are generated from the 
divergent two-electron vertex functions $(59)$ by the singular electron - 
quasiparticle transformation $(58)$ control the charge and spin 
static quantities of the 1D many-electron problem.

On the one hand, the parameters $v_{\rho}^{\iota}$  
and $v_{\sigma_z}^{\iota}$ of Eq. $(59)$ can be shown to be 
fully determined by the two-quasiparticle functions $(60)$.
By inverting relations $(60)$ with the vertices given
by Eq. $(59)$ expressions $(61)$ follow. Physically,
the singular electron - quasiparticle transformation $(58)$
maps the divergent two-electron functions onto the
finite parameters $(60)$ and $(61)$.

On the other hand, the ``velocities'' $(61)$ play a relevant
role in the charge and spin conservation laws and
are simple combinations of the zero-momentum two-pseudoparticle
forward-scattering $f$ functions and amplitudes introduced
in Refs. \cite{Carmelo92} and \cite{Carmelo92b},
respectively. Here we follow Ref. \cite{Carmelo94c}
and use the general parameter $\vartheta $ which refers
to $\vartheta =\rho$ for charge and $\vartheta =\sigma_z$ for 
spin. The interesting quantity associated with the equation of 
motion for the operator ${\hat{\rho}}_{\vartheta }^{(\pm)}(k,t)$ 
defined in Ref. \cite{Carmelo94c} is the following ratio

\begin{equation}
{i\partial_t {\hat{\rho}}_{\vartheta }^{(\pm)}(k,t)\over 
k}|_{k=0} = {[{\hat{\rho}}_{\vartheta }^{(\pm)}(k,t),
:\hat{{\cal H}}:]\over k}|_{k=0} = v_{\vartheta}^{\mp 1}
{\hat{\rho}}_{\vartheta }^{(\mp)}(0,t) \, ,
\end{equation}
where the functions $v_{\vartheta}^{\pm 1}$ $(61)$ 
are closely related to two-pseudoparticle 
forward-scattering quantities as follows

\begin{eqnarray}
v_{\vartheta}^{+1} & = & {1\over
{\left[\sum_{\alpha ,\alpha'}{k_{\vartheta\alpha}k_{\vartheta\alpha'}
\over {v_{\alpha}v_{\alpha '}}}
\left(v_{\alpha}\delta_{\alpha ,\alpha '} - {[A_{\alpha\alpha '}^{1}+
A_{\alpha\alpha '}^{-1}]
\over {2\pi}}\right)\right]}}\nonumber \\ 
& = & {1\over {\left[\sum_{\alpha}{1\over {v_{\alpha}}}
\left(\sum_{\alpha'}k_{\vartheta\alpha '}\xi_{\alpha\alpha 
'}^1\right)^2\right]}} \, ,
\end{eqnarray}
and

\begin{eqnarray}
v_{\vartheta}^{-1} & = &
\sum_{\alpha ,\alpha'}k_{\vartheta\alpha}k_{\vartheta\alpha'}
\left(v_{\alpha}\delta_{\alpha ,\alpha '} + {[f_{\alpha\alpha '}^{1}-
f_{\alpha\alpha '}^{-1}]\over {2\pi}}\right)\nonumber \\ 
& = & \sum_{\alpha}v_{\alpha}
\left(\sum_{\alpha'}k_{\vartheta\alpha '}
\xi_{\alpha\alpha '}^1\right)^2 \, .
\end{eqnarray}
Here $k_{\vartheta\alpha}$ are integers given by $k_{\rho 
c}=k_{\sigma_{z} c}=1$, $k_{\rho s}=0$, and $k_{\sigma_{z} s}=-2$, 
and the parameters $\xi_{\alpha\alpha '}^j$ are defined
in Eq. (A7). In the rhs of Eqs. (D2) and (D3) $v_{\alpha }$
are the $\alpha $ pseudoparticle group velocities (A6), the $f$ functions
are given in Eq. (A4) and $A_{\alpha\alpha'}^{1}=A_{\alpha\alpha'
}(q_{F\alpha}^{(\pm)}, q_{F\alpha'}^{(\pm)})$ and 
$A_{\alpha\alpha'}^{-1}= A_{\alpha\alpha'}(q_{F\alpha}^{(\pm)},
q_{F\alpha'}^{(\mp)})$, where $A_{\alpha\alpha'}(q,q')$ are 
the scattering amplitudes given by Eqs. $(83)-(85)$ of 
Ref. \cite{Carmelo92b}. 

The use of relations $(61)$ and of Eqs. (A5), (A6),
(A8), (D2), and (D3) shows that the parameters $(60)$ and corresponding 
charge and spin velocities $v_{\vartheta}^{\pm 1}$ can also be expressed 
in terms of the pseudoparticle group velocities (A6) and Landau 
parameters (A8). These expressions are given in Eq. $(62)$ 
and in the Table.

The charge and spin velocities control all static quantities 
of the many-electron system. They determine, for example, 
the charge and spin susceptibilities, 

\begin{equation}
K^{\vartheta }={1\over {\pi v_{\vartheta}^{+1}}} \, , 
\end{equation}
and the coherent part of the charge and spin conductivity 
spectrum, $v_{\vartheta}^{-1}\delta (\omega )$, 
respectively \cite{Carmelo92,Carmelo92b,Carmelo94c}. 

%****************************************************************
%********************* R E F E R E N C E S **********************
%****************************************************************

\newpage
\centerline{TABLE}
\vspace{0.10cm}

\begin{tabbing}
\sl \hspace{2cm} \= \sl $v^{\iota}_{\rho }$ 
\hspace{3cm} \= \sl $v^{\iota}_{\sigma_z }$\\
$\iota = -1$ \> $v_c + F^1_{cc}$ \> 
$v_c + F^1_{cc} + 4(v_s + F^1_{ss} - F^1_{cs})$ \\ 
$\iota = 1$ \> $(v_s + F^0_{ss})/L^0$ \> 
$(v_s + F^0_{ss} + 4[v_c + F^0_{cc} + F^0_{cs}])/L^0$ \\ 
\label{tableI}
\end{tabbing}
Table I - Alternative expressions of the parameters $v^{\iota}_{\rho }$ 
(D1)-(D4) and $v^{\iota}_{\sigma_z }$ (D2)-(D5) in terms of 
the pseudoparticle velocities $v_{\alpha}$ (A6) and Landau parameters 
$F^j_{\alpha\alpha '}$ (A8), where $L^0=(v_c+F^0_{cc})
(v_s+F^0_{ss})-(F^0_{cs})^2$.
\end{document}